\documentclass[aip,jcp,%
 amsmath,amssymb,preprint,
,showkeys]{revtex4-1}

\usepackage{graphicx}
\usepackage{dcolumn}
\usepackage{bm}
\begin{document}
\title[]{Collection efficiency of photoelectrons injected into near- and supercritical argon gas}
\author{A. F. Borghesani}\email{armandofrancesco.borghesani@unipd.it.}
\affiliation{ 
 CNISM-Unit, Department of Physics \& Astronomy,  University of Padua\\via F. Marzolo 8, I-35131 Padua, Italy}
\author{P. Lamp}\altaffiliation{Present address: BMW, Munich, Germany}
 \affiliation{
Max-Planck-Institut fuer Physik u. Astrophysik\\ Munich, Germany
}

\date{\today}
\begin{abstract}
Injection of photoelectrons into gaseous or liquid dielectrics  is a widely used technique to produce cold plasmas in weakly ionized systems for investigating the transport properties of electrons. We report measurements of the collection efficiency of photoelectrons injected into dense argon gas for $T=152.7\,$K, close to the critical temperature $T_{c}\approx 150.9\,$K, and for $T=200.0\,$K.
The high-field data agree with the Young-Bradbury model and with previous measurements below $T_{c}$ and at an intermediate temperature above $T_c .$  The effective, density-dependent electron-atom momentum transfer scattering cross section can be deduced.
However, the weak-field data near $T_{c}$ show large deviations from the theoretical model. We 
 show that the electron behavior at weak field is influenced by electrostriction effects that are only important near the critical point.
\end{abstract}

\maketitle

\section{\label{sec:Intro}Introduction}
The study of the injection of charge carriers 
 into a gaseous or fluid dielectric is important for elucidating processes of both technological and fundamental interest whose knowledge is still far to be complete~\cite{schmidtCRC}. 

Electrons can be emitted  in several ways, for instance, using photo-~\cite{schmidt1977,borg1986} or tunnel cathodes~\cite{silver1967,onn1970}. They are  directly injected into the conduction band of the dielectric medium. The energy separation between the Fermi level in the emitter and the bottom of the conduction band of the insulator can be measured~\cite{Blossey1974}. This methodological approach allows  researchers to investigate the energy levels of the electrons in the medium and to shed light on the nature of their states in a dielectric host~\cite{schmidt1977,smejtek1973}. 

The energy, with which electrons are injected into the medium, is typically in excess of thermal. The excess energy is continuously dissipated via scattering events that eventually lead electrons to thermalization. 
Under the action of an externally applied electric field in competition with the image force field, electrons drift towards the anode, at which they are collected. The measured electron mobility is determined by  scattering processes~\cite{Huxley}. 

However, some of the injected electrons can be backscattered via the same scattering mechanisms and can be recaptured by the cathode. Thus, a measure of the electron collection efficiency may give useful pieces of information about the scattering processes a hot electron undergoes on its way to thermalization~\cite{onn1970}. It is expected that mobility-  and collection efficiency measurements give a coherent picture of the scattering processes involving the excess electrons and the atoms of the medium.

Dense rare gases represent a model system of a disordered medium. The possibility of easily varying their density in a wide range between the dilute gas- (with density $N\approx 10^{-2}\,$nm$^{-3}$) and the liquid region ($N\approx 20\,$nm$^{-3}$) allows researchers to study how the electronic state and transport properties depend on density and degree of disorder.

In the past, Young and Bradbury (YB)  developed a simple yet successful model to relate the collection efficiency of photoelectrons injected into a dilute rare gas to the strength of an electric field externally applied to the electrodes~\cite{YB1933}. According to this model, the collection efficiency is determined by the scattering processes and the scattering cross section can be determined from the electric field dependence of the collection efficiency. This model is quite successful at predicting the overall dependence of the collection efficiency on the electric field strength at low gas density though the hypotheses, on which it is based, are somehow unreasonable.

More recently, the injection of electrons in dense argon gas and liquid has been investigated by using thin-film cold-cathode emitters~\cite{smejtek1973}. The analysis of the data according to the YB model in a wide density range has resulted in an unexpected density dependence of the electron-atom momentum transfer scattering cross section. Although the collection efficiency data of Smejtek {\em et al.} agree very well with our more recent measurements~\cite{borg2011,BLIEEE2012}, their determination of the cross section is wrong because the correct description of the multiple scattering (MS) effects, which are very important at high density, was unavailable at the time of their measurements~\cite{bsl1992}.

Numerical simulations based on Monte Carlo (MC) techniques based on classical trajectories have also been carried out in order to statistically investigate the electron injection process as a function of electric field and temperature in a way that is not influenced by the flaws of the YB model~\cite{kuntz1982}. As a matter of fact, MC simulations lead to a correct field dependence of the charge collected at the anode  without giving a physical explanation of the result. Moreover, they fail at predicting the density dependence of the experimental observation because, even in this case, the MS effects are not accounted for at all.

Nowadays, the scattering processes in gases at high density are very well understood~\cite{bsl1992} and the r{\^o}le of MS is clear. The difference with scattering at low density is due to the fact that
multiple scattering affects the electron-atom scattering cross section. The analysis of the drift mobility data has resulted in the formulation of a unified heuristic model, in which MS effects lead to a density dependence of the effective cross section~\cite{bsl1992}. This model agrees well with the mobility data in gases endowed with positive scattering length (such as helium and neon)~\cite{borg1988,borg1990,borg2002} up to densities, at which localization of electrons in density fluctuations sets in~\cite{Hernandez1991}, and agrees with the mobility results in negative length gases (argon) up to even larger densities because electrons still propagate through them as quasifree particles with very long mean free path (mfp)~\cite{bsl1992,Borghesani2001,lekner1967}.

The theoretical tools for the correct interpretation of the charge collection process are thus available. For this reason, within our program aimed at measuring the electron drift mobility in dense rare gases we have also carried out measurements of charge collection efficiency.  

In previous papers we have reported the experimental data in the one-phase region of argon for $T=142.6\,$K below the critical temperature~\cite{borg2011} as well as for $T=177.3\,$K in the supercritical region~\cite{BLIEEE2012}. Those measurements have been analyzed in terms of the YB model by taking into account the MS effect as requested by the model developed for the drift mobility. The main result is that the effective, density-dependent momentum transfer scattering cross section determined from the analysis of the collection efficiency data is absolutely compatible with its determination from the drift mobility measurements and that it can theoretically be calculated using the experimental zero-density electron- atom scattering cross section.

In this paper we present further data at $T=200.0\,$K and $T=152.7\,$K. On one hand, the goal of the measurements is to confirm the prediction of the analysis carried out for the previously investigated temperatures. On the other hand, the temperature $T=152.7\,$K has been chosen because it is so close to the critical temperature $T_{c}=150.9\,$K that very large densities can easily be reached. 
Both series of measurements confirm the results of the YB analysis. 
However, the data at weak field of the lower $T$ isotherm unexpectedly show that another phenomenon, namely electrostriction, determines the collection efficiency of electrons. 

The paper is organized as follows. In Sect.~\ref{sect:ExpDet} the details of the experiment are reported. The experimental data are described in Sect.~\ref{sect:ExpRes}. The discussion of the data in Sect.~\ref{sect:Disc} is divided into two parts. The first one, Sect.~\ref{sect:HF}, describes the analysis of the high-field data according to the YB and the MS models. The second one, Sect.~\ref{sect:lowE}, describes the low-field data and their relationship with the phenomenon of electrostriction.
Finally, the conclusions are drawn.

\section{\label{sect:ExpDet}Experimental details}
\begin{figure}[t!]
\centering
\includegraphics[width=\columnwidth]{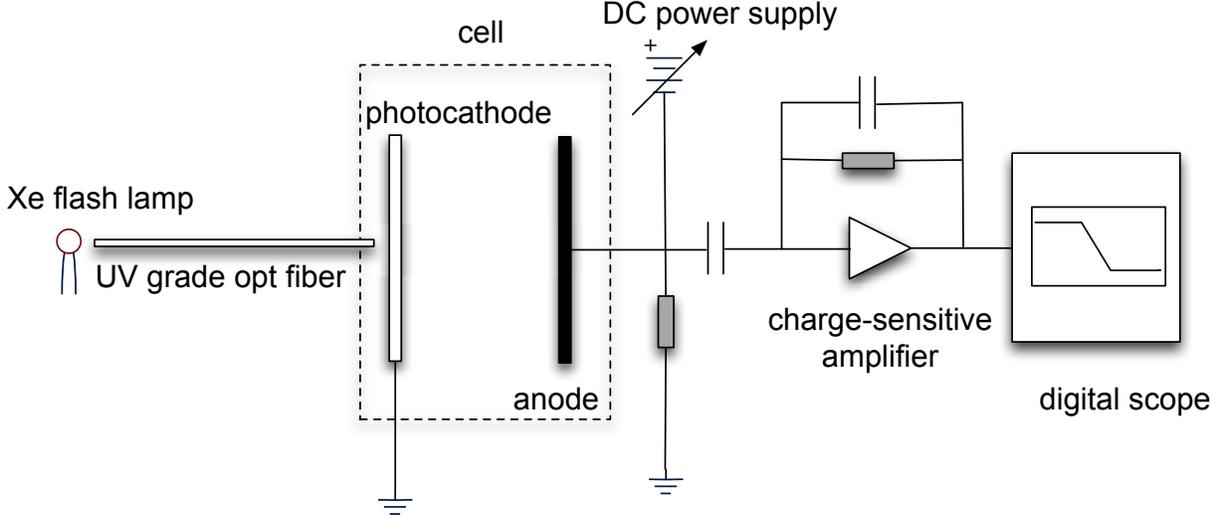}
\caption{\label{fig:apparatus}\small 
Simplified schematics of the experimental apparatus.} 
\end{figure}

Electron injection into the dense gas is accomplished by using the well known pulsed Townsend photoemission method\cite{Huxley}. We previously used this technique for electron mobility measurements in dense neon\cite{borg1988,borg1990}, helium\cite{borg2002}, and argon\cite{bsl1992}. The experimental apparatus, schematically shown in Fig.~\ref{fig:apparatus}, was used for electron mobility measurements in liquid, gaseous, and critical argon and details have been published elsewhere\cite{lamp1989,lamp1990}.
We briefly recall here its most relevant features.

The sample cell is made of a massive copper block that can withstand pressure up to more than $7\,$MPa. Pressure is measured with $\pm 1\,$kPa accuracy. 
 The cell is mounted inside a cryostat and is thermoregulated within 1 mK in the range $80\,\mbox{K}<T<300\,$K. 
 
Argon of the highest commercially available purity (99.9999\% by vol.) is used. Further purification is accomplished by flowing the gas through an Oxisorb filter (Messer Griesheim, Germany)\cite{torzo1990} so as to reduce the impurity content down into the parts per billion range or less as required to carry out accurate electron mobility measurements\cite{bsl1992}. The gas density is computed from temperature and pressure by using an accurate equation of state\cite{Tegeler1997}.

An ultraviolet (UV)-grade quartz window coated with a $\approx 10\,$nm thick Au layer is used as both photocathode and electrode for the drift voltage\cite{borg1986}. The UV light source is a commercially available Xe flash lamp (Hamamatsu, model L2435) delivering light pulses lasting $\approx 1\,\mu$s each. The emitted light spectrum can be described by an asymmetric Gaussian peak centered at a wavelength $\lambda_m \approx 232\,$nm with left and right widths $(-6,+28)\,$nm, respectively. $\lambda_m$ corresponds to photons of $\approx 5.4\,$eV energy.

 The light is guided onto the photocathode by means of a UV-grade quartz fiber. Electrons extracted from the photocathode are injected into the gas and drift toward the anode under the action of an externally applied electric field. The electron current is integrated by an active integrator (charge sensitive amplifier) connected to the anode. Typically, a few millions of electrons per pulse are released {\em in vacuo}. The charge sensitive amplifier response is $\approx 3\,$mV/fC. The resulting output voltage signal is recorded and displayed on a digital storage scope.
 
In order to improve the signal-to-noise ratio, the electronic signals of 256 light bursts are averaged together for each electric field strength settings. The resulting average signal waveform is fetched by a personal computer connected to the scope for offline data processing. The waveforms are thus analyzed with numerical techniques to determine signal amplitude, hence collected charge, and electron drift time\cite{borg1990b}.

\section{\label{sect:ExpRes}Experimental results}
We report the experimental results of the measurements of the charge collected at the anode in dense argon gas in the supercritical region 
 at $T=152.7\,$K, quite close to the critical temperature, $T_c \approx 150.9\,$K, and 
 at the much higher temperature $T=200.0\,$K. The two temperatures are sufficiently different so as to result in a 30 \% difference in the electron thermal energy. Moreover, the lower temperature is close enough to $T_c$ so as to allow us to investigate the electron behavior at very high densities without having to resort to exceedingly high pressures. Actually, a maximum number density $N\approx 11.4\,\mbox{nm}^{-3}= 1.4N_c$ has been attained, where the critical density is $N_c\approx 8.08\,$nm$^{-3}.$

Experimental data at different temperatures in the one-phase region for $T=142.6\,\mbox{K}<T_c$\cite{borg2011} and in the supercritical region for $T=177.3\,$K\cite{BLIEEE2012} have been previously published.

For all measurements, the electric field $E$ is set by the D.C. power supply in the range $0.1\,\mbox{kV/m}< E<40\,\mbox{kV/m}.$ At the high end of its range, it is weak enough to avoid breakdown or ionization of the gas. At the low end of the range we have always checked that contact potential effects are absent by realizing that the measured electron mobility is field independent~\cite{bsl1992,Borghesani2001}.

It is customary to choose $E/N$ as a parameter to rationalize the data in swarm experiments. Actually, the drifting electrons scatter off the gas atoms and the mean energy in excess of thermal gained from the field over a mean free path $\ell=(1/N\sigma_\mathrm{mt})$ is $eE\ell,$ where $\sigma_\mathrm{mt}$ is the electron-atom momentum transfer scattering cross section. Thus, for a given (and constant) cross section, $E/N$ is proportional to the excess electron energy gained from the field.

\begin{figure}\centering
\includegraphics[width=\columnwidth]{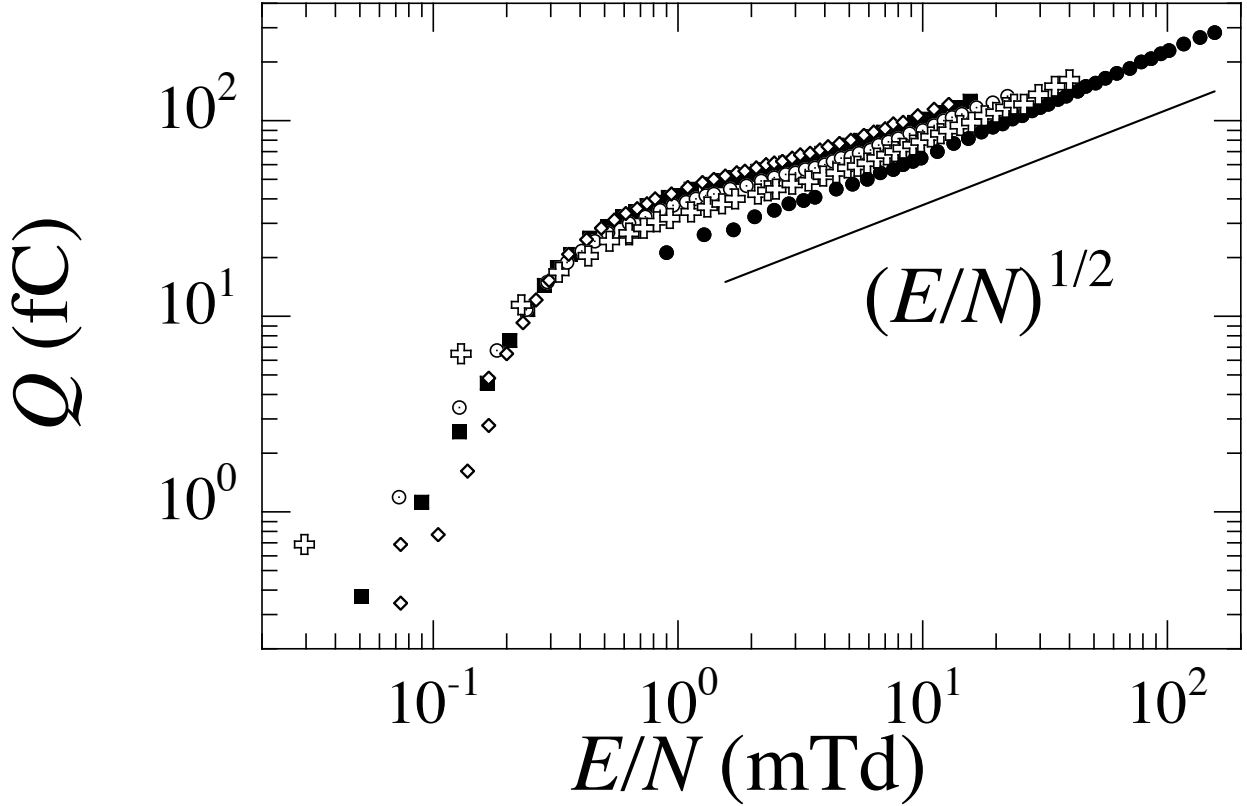}
\caption{\label{fig:Q200ENACFIL}\small 
$Q$ vs $E/N$ at $T=200\,$K for some isopycnals of density $N \,(\mbox{nm}^{-3}) =3.31$ (open diamonds), 2.57 (closed squares), 1.80 (open circles), 1.01 (crosses), and 0.256 (closed circles) Only a few isopycnals are shown for the sake of clarity. The solid line represents the $(E/N)^{1/2}$-dependence.} 
\end{figure}

For this reason, in Fig.~\ref{fig:Q200ENACFIL} we report the data of the charge $Q$ collected at the anode in supercritical argon gas for $T=200.0\,$K as a function of the reduced electric field $E/N$ for several isopycnals. The density range investigated for this temperature is $0.26\,\mbox{nm}^{-3}<N<3.31\,\mbox{nm}^{-3},$  whereas the reduced electric field range is $10^{-2}\,\mbox{mTd}<E/N<160\,\mbox{mTd} $  ($1\,$Td$=10^{-21}\,$V$\,\mbox{m}^2).$

The behavior of $Q$ as a function of $E/N$ for $T=200\,$K favorably compares with the previously measured data for $T=142.6\,$K~\cite{borg2011} and for $T=177.3\,$K~\cite{BLIEEE2012}. For $E/N\gtrsim 1.5\,$mTd, $Q\propto (E/N)^{1/2}$ for all densities. In this region, the charge $Q$ collected at fixed $E/N$ increases with increasing the gas density $N.$ A similar behavior, both as a function of $E/N$ and of $N,$ was also observed for $T=160\,$K~\cite{smejtek1973}, though for $E/N>10\,$mTd, i.e., in a range of field strengths much higher than in the present case. By contrast, Monte Carlo simulations~\cite{kuntz1982}, that reproduce the correct reduced field dependence, predict a density ordering of the data opposite to the experiment. 

For weaker fields, the density ordering of the data is reversed and also the field dependence changes towards a $(E/N)^{\gamma}$-law  with $\gamma\geq 1/2$ and approaching 1 for the highest densities. 

It is possible to rule out any space-charge effect as the cause of the observed low-field behavior of $Q$ because the the amount of charge injected during each light pulse is quite small and because the zero-field electron mobility $\mu\approx 0.1 \,$m$^2\,$V$^{-1}\,$s$^{-1}$ is large even at the highest densities~\cite{bsl1992,Borghesani2001}. Moreover, the residual content of O$_2$ impurities is so small that the concentration of slow O$_2^-$ ions is negligible. 

Such a conclusion is supported by the analysis of the behavior of the photoelectric current measured at constant $N$ in liquid argon at the normal boiling point $T\approx 87\,$K as a function of the electric field $E$~\cite{sakai1987,sakai1991}. At low injection level, the photocurrent rises $\propto E$ at first and shows a crossover to a $E^{1/2}$-law at a field that roughly corresponds to $1\,$mTd. As the injection level is raised, the field dependence at low field gradually changes toward a $E^2$-law, typical of a space-charge-limited regime, and the crossover region shifts to increasingly larger field strength.

\begin{figure}
\centering
\includegraphics[width=\columnwidth]{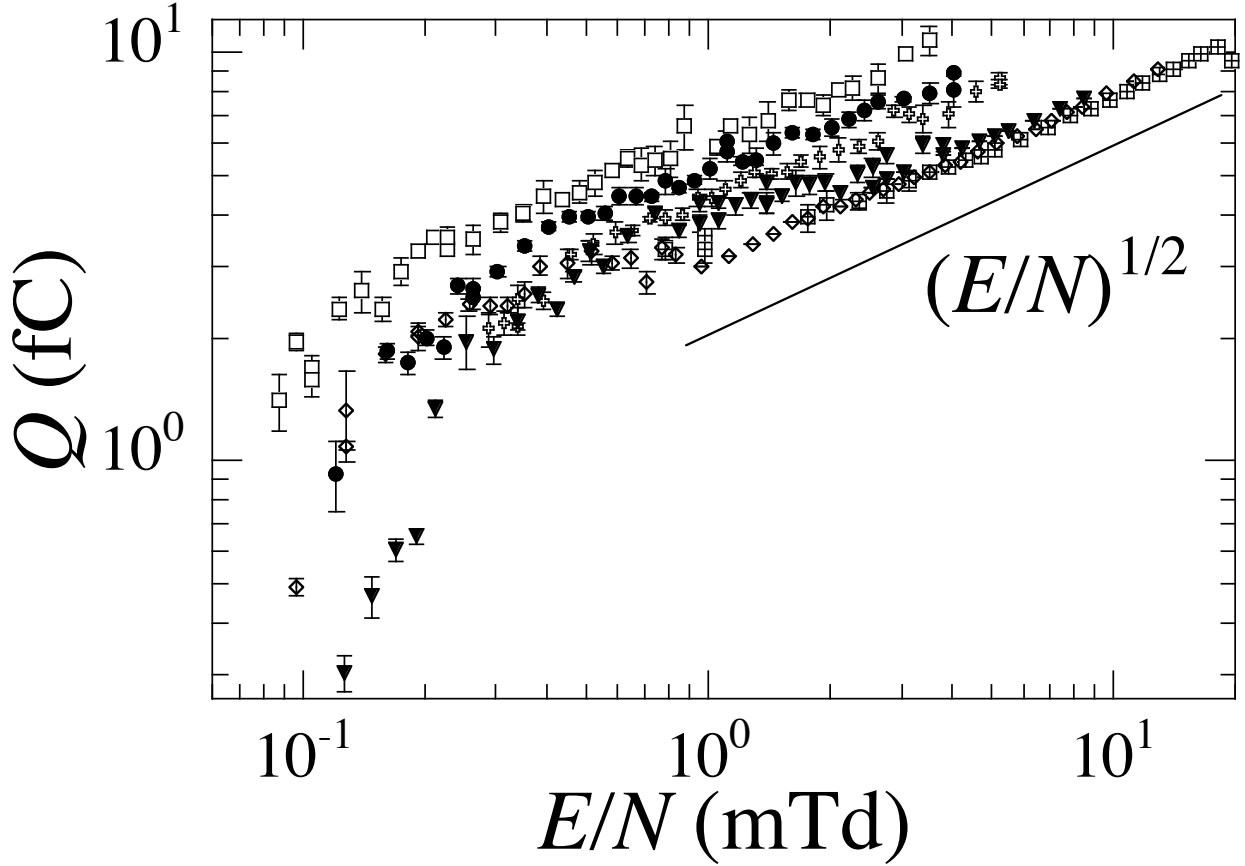}
\caption{\label{fig:QEN153KAEJPSX}\small 
$Q$ vs $E/N$ at $T=152.7\,$K for some isopycnals of density $N \,(\mbox{nm}^{-3}) =11.39$ (open squares), 9.89 (closed circles), 7.63 (crosses), 4.71 (triangles), 3.11 (diamonds), and 0.5 (crossed squares). Only a few isopycnals are shown for the sake of clarity. Solid line: $(E/N)^{1/2}$-law.}
\end{figure}

The data obtained for $T=152.7\,$K are displayed in Fig.~\ref{fig:QEN153KAEJPSX}. 
We first note that the collected charge is approximately a factor 10 smaller than for the experiment at $T=200\,$K. This is due to the fact that the UV-light source intensity was reduced by a similar factor in order to minimize the heat input into the gas when working close to the critical temperature. As a result the statistical quality of the data is a bit worse than for the other temperature.

\begin{figure}[t!]
\centering
\includegraphics[width=\columnwidth]{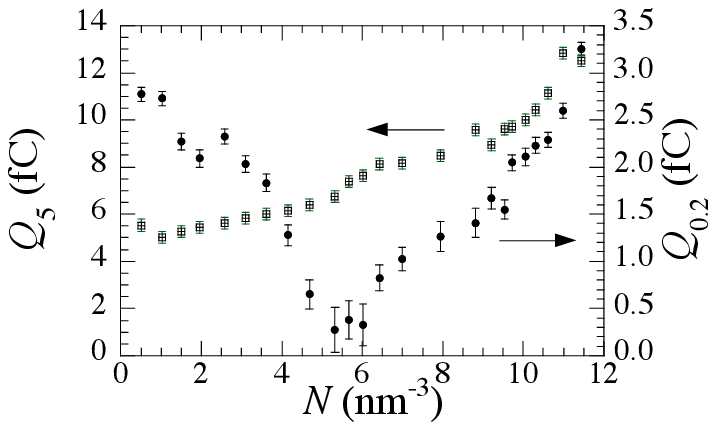}
\caption{\label{fig:QNT15302and5mTd} \small $Q$ vs $N$ at $T=152.7\,$K for fixed values of $E/N.$ 
$Q_{\mathrm{5}}$ charge collected for $E/N=5\,$mTd (crossed squares, left scale). $Q_{\mathrm{0.2}}$ charge collected for $E/N=0.2\,$mTd (circles, right scale). } 
\end{figure}

In spite of this, the data for $T=152.7\,$K show the same behavior of all other sets of measurements at high field strengths, i.e., $Q\propto (E/N)^{1/2},$ where the density ordering of the data is normal, i.e., $Q$ increases with increasing $N$ at constant $E/N.$ It also shows a crossover region about $0.5<E/N< 1 \,$mTd towards a steeper dependence on $E$ at weaker fields. 

However, in the weak field region the density ordering of the data is not simply reversed as for the other temperatures. Actually, $Q$ shows a very peculiar dependence on $N$ in the weak field region. This particular behavior can be best realized by plotting the data collected at constant $E/N$ in the two extreme regions of weak- and strong fields as a function of $N,$ as can be seen in Fig.\ref{fig:QNT15302and5mTd}.

$Q_{5}$ is the charge collected at the anode for $E/N=5\,$mTd  and $Q_{0.2}$ is the charge collected for $E/N=0.2\,$mTd.
At high fields, $Q_{5}$ increases with increasing $N,$ as expected on the basis of the results of the measurements at all temperatures~\cite{borg2011,BLIEEE2012}. On the contrary, at weak fields, $Q_{0.2}$ initially decreases with increasing $N.$ It shows a minimum at a density $N\approx 6\,$nm$^{-3} $ well below the critical density $N_{c}=8.08\,$nm$^{-3}.$ Finally, $Q_{0.2}$ starts increasing again with $N.$

We will show that the peculiar behavior of $Q$ vs $N$ at low field is due to electrostriction effects that are particularly important near the critical point whereas the data at higher fields are not influenced by electrostriction.

For this reason, the analysis of the experiment will be carried out in two distinct parts. In the first one, we will discuss the high-field data that give pieces of information on the electron-atom momentum transfer scattering cross section and the present results will be compared with literature results. 
 
 In the second part, the weak-field data will be discussed in the context  of the electrostriction model developed earlier in order to rationalize the experimental results of the mobility of slow O$_{2}^{-}$ ions in near critical argon gas~\cite{cpl}.

\section{\label{sect:Disc}Discussion}
In this section we discuss the present experimental data and compare them with previous literature results.
As anticipated, we first discuss the high-field data because they can be analyzed within the theoretical YB model and because they can be directly compared with previous measurements of charge collection efficiency in dense argon gas.

In the second part, we will discuss the  low-field data because their behavior is deeply influenced by the fact that the measurements are carried out very close to the critical temperature. They cannot be rationalized without invoking the effect of criticality.

\subsection{\label{sect:HF}The high-field data}
In this section, we show how the YB model, if supplemented with the knowledge of multiple scattering effects gathered in the experiments of electron mobility in dense noble gases~\cite{borg1988,borg1990,bsl1992,Borghesani2001}, can successfully be applied to get an independent determination of the electron-atom momentum transfer scattering cross section~\cite{borg2011, BLIEEE2012}. To this goal, it is convenient to describe the fundamentals of the electron photoemission and thermalization processes, the details of the YB model, and the main features of the multiple scattering (MS) model.

\subsubsection{\label{sec:PhPr}The photoemission and thermalization processes}

\begin{figure}[b!]
\centering
\includegraphics[width=\columnwidth]{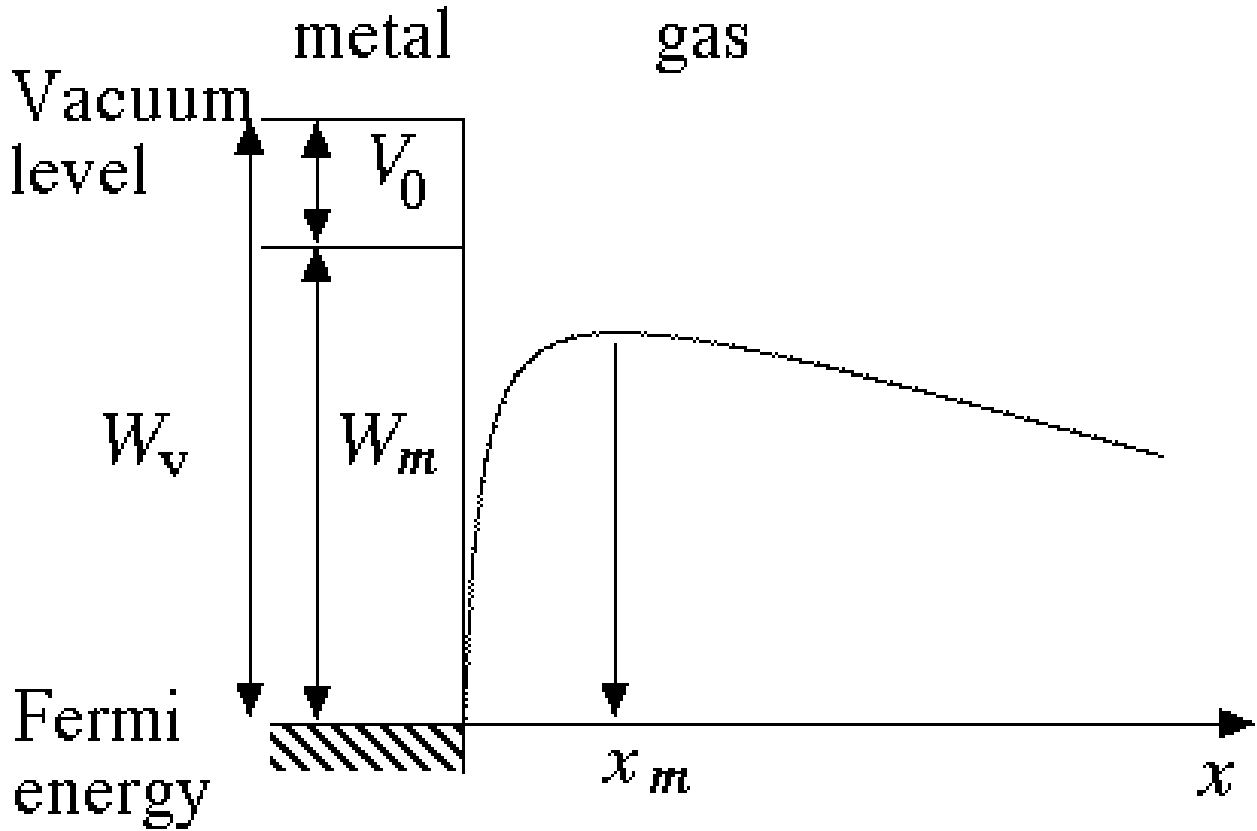}
\caption{\label{fig:V0} \small  Schematic diagram of the energy band at the metal-gas interface.}
\end{figure}

Electrons are extracted from a metal photocathode into vacuum by UV light when the photon energy exceeds the work function $W_{v}$ of the metal. A diagram of the energy band at the metal-gas interface is shown in Fig.\ref{fig:V0}.
If the photocathode is immersed in a dielectric medium, the threshold energy for photoemission $W_{m}$ is changed with respect to the vacuum value $W_{v}$ by the amount $V_{0}(N)$
\begin{equation}
W_{m}= W_{v} + V_{0}(N)
\label{eq:V0}\end{equation}
$V_{0}(N)$ is considered as the bottom energy of the electron conduction band in the medium~\cite{schmidt1977} and depends on the density $N.$ $V_{0}$ may be positive or negative depending on the relative weight of the positive electron excess kinetic energy due to the shrinking of the volume available to free electrons when the density is increased, and the negative polarization energy due to the medium polarization induced by electrons~\cite{SJC}.
 In the case of argon, $V_{0}<0$ and less photon energy than in {\em vacuo} is required to extract photoelectrons from the metal~\cite{reininger1983,bachelet1986}.
 
 Electrons photoemitted into the medium show a quite broad energy distribution~\cite{Fowler1931,dubridge1932,dubridge1933} whose highest energy is $\mathcal{E}_{0} =(hc/\lambda_{m})-W_{m},$ where $\lambda_{m}$ is the shortest wavelength in the UV light used and $h$ and $c$ are the Planck's constant and the speed of light in {\em vacuo}, respectively.
 
 The just injected electrons are epithermal and drift through the gas under the combined action of diffusion, of their own image field that pulls them back to the cathode~\cite{jackson}, and of the externally applied electric field $E$ that pushes them towards the anode. The net potential energy is 
 \begin{equation}
 V(x)= -\frac{1}{4} \frac{e^{2}}{4\pi\epsilon_{0}Kx} -eEx
 \label{eq:vx}\end{equation}
 in which $x$ is the distance from the cathode into the gas, $\epsilon_{0}$ is the vacuum permittivity, and $K$ is the relative dielectric constant of the medium. At the densities of the present experiment, $K\approx 1$ within a few \%~\cite{maitland}. 
 
 $V(x)$ has a maximum $V_{m}= -2eEx_{m}$ at the distance $x_{m}=(e/16\pi\epsilon_{0}KE)^{1/2.} $ The action of the electric field $E$  is thus to lower the threshold energy for injection by the Schottky correction $\Delta W= \vert V_{m}\vert .$ In the present experiment, 
 at the boundaries of the electric field strength range, we obtain $x_{m}\approx 2\,\mu$m and $V_{m}\approx 40\,\mu$eV for $E=0.1 \,$kV/m, and $x_{m}\approx 95\,$nm and $V_{m}\approx  8\,$meV for $E=400\,$kV/m.
We conclude that, in the present experiment, the electric field strength is always so small as to yield a Schotty barrier lowering smaller than the electron thermal energy $\mathcal{E}_{T}=(3/2) k_{\mathrm B}T =20\,$ meV for $T=152.7\,$K and $\mathcal{E}_{T}=26\,$meV for $T=200\,$K. Moreover, the position of the potential energy maximum is located very deep into the medium, even on the mfp scale. 

Once injected into the medium with excess energy $\mathcal{E}_{0},$ the epithermal electrons undergo scattering processes leading them to backdiffusion and thermalization. The probability for electrons to be collected at the anode mainly depends on the length of the path travelled before getting thermalized compared with $x_{m}$~\cite{onn1969}. In pure argon gas, only momentum exchange collisions occur that randomize the electron velocities leading to a slow loss of the initial kinetic energy of the electrons. A first possibility for the electrons is to be immediately backscattered upon injection and to get back to the cathode~\cite{onn1969}. Once this happens, the chance that a backscattered electron still diffuse towards the anode over the potential barrier or directly  tunnel through it is negligible because $x_m$ is large, because the electric field strength is quite weak, and because the temperature is quite low~\cite{Blossey1974}.  

The remaining electrons that are not backscattered keep diffusing and lose their excess energy in a huge number of scattering events until they thermalize beyond the potential energy maximum and are collected to the anode~\cite{smejtek1973,onn1969}. 

The actual physical situation, however, may be far more complicated than this idealized picture. Actually, the escape probability is smaller the higher is the initial excess electron energy because electrons that have already crossed the the barrier at $x_m$ may have still residual energy to surmount it and any collision able to reverse their motion can send them back to the cathode~\cite{allen1984}.

\subsubsection{\label{sec:YB}The Young-Bradbury model}
In the past many researchers aimed at explaining the ratio of the current measured in presence of a gaseous or liquid medium to the saturation current, i.e., the current collected in {\em vacuo}~\cite{thomson1928,loeb1955,bekirian1968,sakai1991}.
Those results, however, cannot be validated as an exact analytical solution of the Boltzmann's transport equation is not at hand, though numerical MC simulations have been employed to that goal~\cite{kuntz1982}.

Actually, the thermalization process is universally acknowledged to depend on the features of the electron-atom momentum transfer scattering cross section but is very complicated because a huge number of collisions is involved~\cite{mozumder1980,kuntz1982}. Nonetheless,  the extremely simplified YB model~\cite{YB1933} does a good job at describing the experimental results of charge collection efficiency in low density gases. The simplifying assumptions of the model and the results it provides are subject of strong criticism~\cite{smejtek1973,kuntz1982} but, still, its simplicity is appealing. We will show that the YB model, if enhanced by the knowledge about multiple scattering effects gathered from the electron mobility measurements in dense noble gases, can still provide a very accurate description of the experimental data.

According to the model, an electron is removed from the current stream as soon as it is scattered backwards at such an angle that it can reach the emitter again. The effect of the image field is neglected but it is clear that it can only ease the process. The model further assumes that only reflections of the electrons at their first encounter are sufficient to calculate the return current.

An electron that undergoes a backscattering event at a distance $x$ from the cathode can return to the emitter if its velocity towards the cathode after the collision, $u,$ is such that 
$(1/2) mu^2\geq eEx,$ i.e., if the kinetic energy associated with the motion towards the cathode is greater of the work done by the field over the distance $x.$
The total electron kinetic energy at a collision with velocity $w$ is $(1/2) mw^2= \mathcal{E}_0 +eEx,$ in which $\mathcal{E}_0$ is the injection energy. Thus, the electron can return to the cathode if it is scattered within a cone subtending the solid angle
\begin{equation}
    \Omega=2\pi \left[
    1- \left(
    \frac{x}{x+\mathcal{E}_0/eE}
    \right)^{1/2}
    \right]
     \label{eq:cone}
\end{equation}
   and the return probability $\mathcal{R}$ is simply defined on a geometrical basis as
   \begin{equation}
       \mathcal{R}(x)=\frac{\Omega}{4\pi}= \frac{1}{2}\left[
    1- \left(
    \frac{x}{x+\mathcal{E}_0/eE}
    \right)^{1/2}
    \right]
    \label{eq:refcoef}
   \end{equation}
If some further hypotheses are assumed, namely, that the electric field is weak enough not to significantly deflect electrons before their first encounter and that the mean free path $\ell$ is negligible compared with the drift distance, the ratio of the observed current $I$ to the saturation current $I_{0}$ or, in the case of pulsed injection, the ratio of the charge $Q$ collected in the gas to the charge $Q_{0}$ collected in {\em vacuo}, can be calculated as
\begin{eqnarray}
\frac{I}{I_{0}}\equiv \frac{Q}{Q_{0}}&=&\int\limits_{0}^{\infty}\ell^{-1}\exp({-x/\ell)} \left[
    \frac{x}{x+\mathcal{E}_0/eE}
    \right]^{1/2}\, \mathrm{d}x
\nonumber \\
&=&\int\limits_{0}^{\infty} e^{-y}\left[ \frac{y}{y+d^{-2}}
\right]^{1/2}\,\mathrm{d}y
\label{eq:QsuQ0}\end{eqnarray}
Here, $d^{2}=eE\ell/\mathcal{E}_{0}=(eE/\mathcal{E}_{0}N\sigma_\mathrm{mt}).$

\begin{figure}[b!]\centering
\includegraphics[width=\columnwidth]{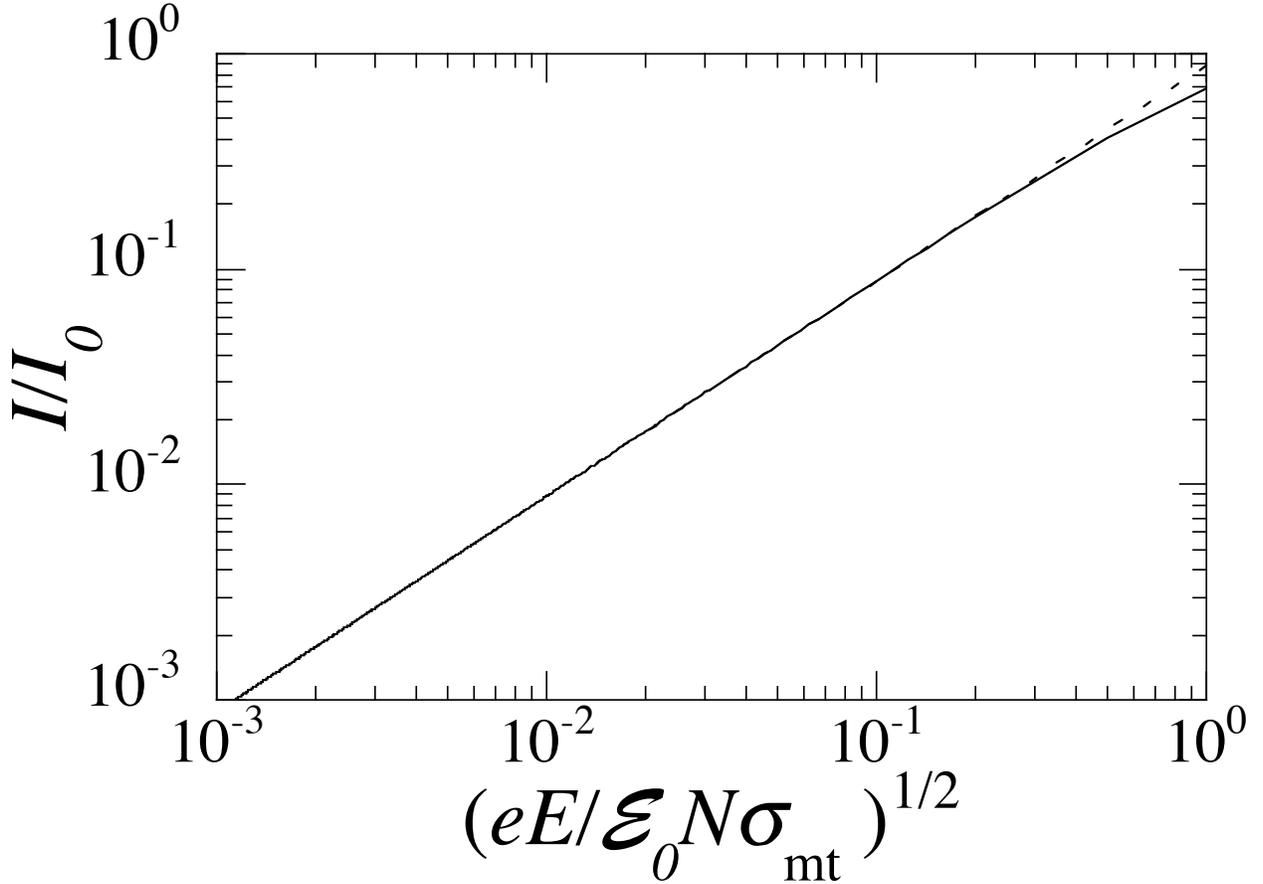}
\caption{\label{fig:Yb} \small Ratio of the collected current  to the saturation current $I/I_0.$ Solid line: Eq.~\ref{eq:QsuQ0}. Dashed line: $\Gamma(3/2)(eE/\mathcal{E}_{0}N\sigma_\mathrm{mt})^{1/2}.$}
\end{figure}

The integral in Eq.~\ref{eq:QsuQ0} is easy to be numerically evaluated as a function of the parameter $d\propto(E/N)^{1/2}$ and the result is shown in Fig.~\ref{fig:Yb}.
For $d= (eE/\mathcal{E}_{0}N\sigma_\mathrm{mt})\lesssim 0.2,$ the ratio $Q/Q_0$ can be very well approximated by 
\begin{equation}
    \label{eq:QQ0}
    \frac{Q}{Q_0}\simeq \Gamma(3/2)\left(\frac{e}{\mathcal{E}_0 \sigma_\mathrm{mt}}
    \right)^{1/2}\left(\frac{E}{N}\right)^{1/2}
\end{equation}
with $\Gamma(3/2)\approx 0.886.$ In the conditions of the present experiment, we estimate $\mathcal{E}_0\approx 0.35\,$eV and assume $\sigma_\mathrm{mt}\approx 8\times 10^{-20}\,$m$^2$ for thermal electrons~\cite{wey1988,haddad1982}, so that we can assert the validity of Eq.~\ref{eq:QQ0} for $E/N$ up to $1\,$Td or even more.

A further condition that must be fulfilled for Eq.~\ref{eq:QsuQ0} to be valid is that the fraction $\mathcal{R}$ of electrons that are backscattered to the cathode after they crossed a distance $\ell$ is smaller than the fraction $\mathcal{T}=1-\mathcal{R}$ of electrons arriving at the anode. How smaller $R$ has to be than $T$ is matter of speculation~\cite{smejtek1973}. We can write
\begin{equation}
{T}(\ell)-\mathcal{R}(\ell) =\left(\frac{eE\ell}{\mathcal{E}_0 +eE\ell}\right)^{1/2}<\beta
\label{eq:tmenor}
\end{equation}
Necessarily, $0<\beta<1.$ In the present experiment, the strongest field is $E=40\,$kV/m, the longest 
mfp can be estimated to be $\ell\approx 10\,$nm, so that the maximum work spent by the field turns out to be
$eE\ell \approx 0.4\,$meV $\ll\mathcal{E}_0\approx 0.35\,$eV and Eq.\ref{eq:tmenor} can be rewritten so as to yield a condition on $E/N$
\begin{equation}
\left(\frac{E}{N}\right)\gtrsim \beta^2\frac{\mathcal{E}_0\sigma_\mathrm{mt}}{e}
\label{eq:hen}
\end{equation}
If we use the value $\beta =0.1$ as suggested in literature~\cite{smejtek1973} and if the previously estimated values of $\mathcal{E}_0$ and of $\sigma_\mathrm{mt}$ are used, Eq.~\ref{eq:hen} yields $E/N\gtrsim 0.3\,$Td. Combining this estimate and the previous one, the validity of the YB model should be limited to the range $0.3\,\mathrm{Td}\lesssim E/N\lesssim 1\,\mathrm{Td.}$ A quick inspection of Fig.~\ref{fig:Q200ENACFIL} and of Fig.~\ref{fig:QEN153KAEJPSX} reveals that the $(E/N)^{1/2}$-law is obeyed over a far wider reduced field range. Previous high-field data~\cite{smejtek1973} as well as low-field data~\cite{borg2011,BLIEEE2012} confirm that $(E/N)^{1/2}$-behavior is shown for much weaker reduced field than predicted by Eq.~\ref{eq:hen}. An obvious reason for this discrepancy may be the choice of too large a value of $\beta.$ At the same time, we have to stress the fact that it is not immediately clear what value for $\sigma_\mathrm{mt}$ should be used in Eq.~\ref{eq:hen} owing to the strong dependence of $\sigma_\mathrm{mt}$ on the electron energy~\cite{haddad1982,wey1988} and to the influence, as will be discussed later, of multiple scattering effects~\cite{bsl1992}.

The data presented here in Fig.~\ref{fig:Q200ENACFIL} and Fig.~\ref{fig:QEN153KAEJPSX} span a low- to intermediate reduced field strength range $2\times 10^{-4}\,\mathrm{Td}<E/N<2\times 10^{-1}\,\mathrm{Td},$ roughly the same range of our previous measurements\cite{borg2011,BLIEEE2012}, and partially overlap old higher-field measurements in the range $5\times 10^{-3}\,\mathrm{Td}<E/N<40\,\mathrm{Td}$~\cite{smejtek1973}. From our data, it can be seen that the validity of the YB model, as expressed by the $(E/N)^{1/2}$-law, extends further down for $E/N\gtrsim 1\,\mathrm{mTd}$ for all densities up to the highest one for $N\approx 11.4 \,$nm$^{-3}$ at $T=152.7\,$K. However, this observation suggests that the hypothesis leading to Eq.~\ref{eq:tmenor} with $\beta \approx0.1$ is quite arbitrary and not very realistic. 

A further issue of discrepancy between the data and the YB model is that the experimental data show a weak upward curvature even in the range in which the $(E/N)^{1/2}$-law should be followed. However, it is very plausible that this deviation from the YB-law occurs because of the energy dependence of $\sigma_\mathrm{mt}$ that increases with energy beyond the Ramsauer-Townsend (RT) minimum.

\subsubsection{\label{sec:CS}Determination of the scattering cross section}
If one assumes that the YB model is correct, Eq.~\ref{eq:QQ0} can be used to determine the momentum-transfer scattering cross section as a function of the gas density $N.$ 

This is a very important issue. Actually, the electron-atom momentum-transfer scattering cross section is either known via theoretical~\cite{haddad1982} or experimental~\cite{wey1988} studies in the limit of $N\rightarrow 0.$ However, over the years, a large number of electron swarm experiments~\cite{levine1967,borg1988,Borghesani1985,Bartels1973,bsl1992}, in which the electron mobility has been measured in noble gases at high density, 
have shown that the cross section is strongly influenced by density effects (namely, multiple scattering effects) and largely deviates from the prediction of kinetic theory~\cite{Huxley}. 

A heuristic model~\cite{bsl1992} has succesfully been developed in order to rationalize those effects in a unified picture. Thus, an independent determination of the cross section from the collection efficiency experiment is important to validate that model.  

The value of the  cross section can be deduced from Eq.~\ref{eq:QQ0} for each value of $N.$
Let us rewrite Eq.~\ref{eq:QQ0} in the form
\begin{equation}
\frac{Q}{Q_0}= A B(N) \left(\frac{E}{N}
\right)^{1/2}
\label{eq:QQ0bis}
\end{equation}
in which $A= e^{1/2}\Gamma(3/2)$ and $B(N)= \left(\mathcal{E}_0\sigma_\mathrm{mt}\right)^{-1/2}.$

It is, thus, possible to calculate the density dependent cross section as
\begin{equation}
\sigma (N) = \frac{B^2(N_0)\mathcal{E}_0(N_0)}{B^2(N)\mathcal{E}_0(N)}\sigma_0
\label{eq:sigmanorm}
\end{equation}
in which $N_0$ is a reference density, $\sigma_0$ is the cross section value from mobility experiments\cite{bsl1992} at the reference density, and $\mathcal{E}_0(N)$ is the density-dependent injection energy. This same same procedural scheme has been used previously~\cite{smejtek1973,borg2011,BLIEEE2012}.

Intentionally, we have replaced the quantity $\sigma_\mathrm{mt}$ with the simpler symbol $\sigma$ because we reserve the former term for the energy-dependent gas-phase electron-atom momentum transfer scattering cross section determined in low-density swarm or crossed-beams experiments, whereas the latter means the density dependent cross section determined from the collection efficiency data. It will later be shown that $\sigma(N)$ can be evaluated from $\sigma_\mathrm{mt}$ if the multiple scattering effects are taken into account.

The slope values $B(N)$ have been obtained by fitting the experimental data for $E/N\gtrsim 1 \,$mTd to the square-root law $(E/N)^{1/2} . $ The injection energy at $N=0$ can be estimated to be $\mathcal{E}_0 (0)\approx 0.356\,$eV. $\mathcal{E}_0(N)$ changes with $N$ because the metal work function is affected by $V_0(N).$
By taking into account Eq.~\ref{eq:V0}, we get
\begin{equation}
\mathcal{E}_0
(N)=\mathcal{E}_0(0)-V_0
(N)\label{eq:e0n}
\end{equation} 
The energy at the bottom of the conduction band is calculated by using experimental data~\cite{reininger1983}, corrected by impurity effects at low density~\cite{Borghesani1991}. They are well interpolated by the formula
\begin{eqnarray}
V_0 (N)&=&V_0(N_1) +a(N-N_1) +\nonumber \\
&&+ \frac{b}{c}\ln\left\{\cosh\left[c\left(N-N_1\right)\right]
\right\}
\label{eq:v0R}
\end{eqnarray}
with $N_1=11.03\,$nm$^{-3},$ $V_0(N_1)=-0.253\,$eV, $a=-3.34\times 10^{-3}\,$eV/nm$^3,$ $b=2.48\times 10^{-2}\,$eV/nm$^3,$ and $c=-0.3\,$nm$^{-3}.$

The contribution of $V_0(N)$ to $\mathcal{E}_0(N)$ is quite substantial in our experiment because the density can be high. For instance, for the isotherm close to the critical itemperature, in which the highest density of $11.4\,$nm$^{-3}$ has been reached, $V_0$ is nearly as large as $0.9\,\mathcal{E}_0 (0)$ whereas even for only $N\approx 3\,$nm$^{-3}$ it amounts to $\approx 25\,\%$ of $\mathcal{E}_0(0).$

For the reference density and cross section $N_{0}$ and $\sigma_{0},$ respectively, we have chosen
$N_{0}=0.769\,$nm$^{-3}$ with $\sigma_{0}\approx 2.33\times 10^{-20}\,$m$^2$ for $T=200\,$K, and $N_{0}=5.36\, $nm$^{-3}$ with $\sigma_{0}\approx 1.01\times 10^{-20}\,$m$^{2}$ for $T=152.7\,$K. These values are deduced from mobility data published elsewhere~\cite{Borghesani2001,Borghesani2003}.

\begin{figure}[b!]\centering
\includegraphics[width=\columnwidth]{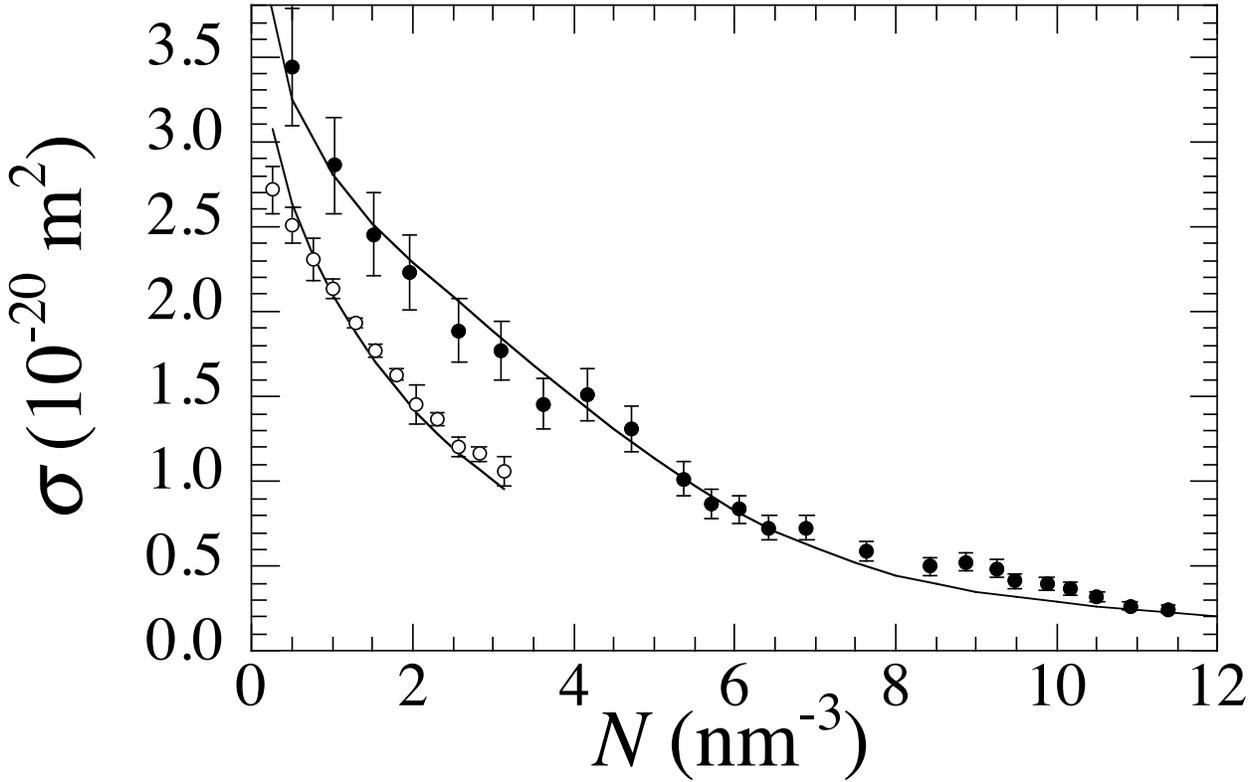}
\caption{\label{fig:SigmaQT153and200} \small  $\sigma$ vs $N$ for $T=200\, $K (open circles) and for $T=152.7\,$K (closed circles) determined from the analysis of $Q$ with the YB model. Solid lines: theoretical prediction of the multiple scattering model.}
\end{figure}

In Fig.~\ref{fig:SigmaQT153and200}
we show the density dependent cross section for the two investigated temperatures $T=200\,$K and $T=1521.7\,$K.

The present data compare favorably with the previously published data at $T=142.6\,$K and $T=177.3\,$K~\cite{borg2011,BLIEEE2012}.  The old data for $T=160\,$K~\cite{smejtek1973} are off by a factor 2 and cannot be shown in this figure. However, as thoroughly discussed  in a previous paper~\cite{borg2011},  Smejteks's data are spoiled by the lack of accurate mobility data and of a sound theoretical model for their interpretation. Actually, it has been shown that the cross section determined from the old charge collection efficiency data~\cite{smejtek1973} are reconciled with the present ones if they are analyzed in the more correct way we are now using.

The density dependence of the electron-atom momentum-transfer scattering cross section determined from the collection efficiency data and shown in Fig.~\ref{fig:SigmaQT153and200} can be explained if the YB model is supplemented with the results of the heuristic model we developed~\cite{borg1988,bsl1992} in order to treat the multiple scattering effects that influence the electron mobility in dense noble gases and that are responsible for the large deviations of the mobility in such systems from the prediction of the kinetic theory~\cite{Huxley}.

\subsubsection{\label{sec:MS}The multiple scattering model}
In a dense gas at fairly low temperatures, in particular in the condition of the present experiment, the electron de Broglie wavelength, its mfp, and the average interatomic distance are comparable to each other. In this situation, multiple scattering effects~\cite{lax1951,foldy1945} cannot be neglected any longer. 

Of the three main MS effects, whose influence has been ascertained in dense noble gases~\cite{bsl1992}, one is particularly important in the present case. The ground state energy of a quasifree electron immersed in a medium is changed with respect to the thermal value by a density dependent, quantum shift~\cite{fermi1934} that is to be considered as the energy at the bottom of the conduction band in the medium, $V_0(N).$

 It is customary~\cite{SJC} to split $V_0 (N)$ in the sum of a potential energy contribution $U_P (N)$  plus a kinetic energy one $E_K (N)$
\begin{equation}
V_0 (N) = U_P (N)+ E_K (N)
\label{eq:Voupek}
\end{equation}
The potential energy term $U_P(N)<0$ stems from the screened polarization interaction between the electron and the surrounding atoms. The kinetic energy term is a quantum, excluded-volume contribution $E_K (N)>0$ arising from the reduction of the volume the electron can access as the gas density is increased.

\begin{figure}
\centering
\includegraphics[width=\columnwidth]{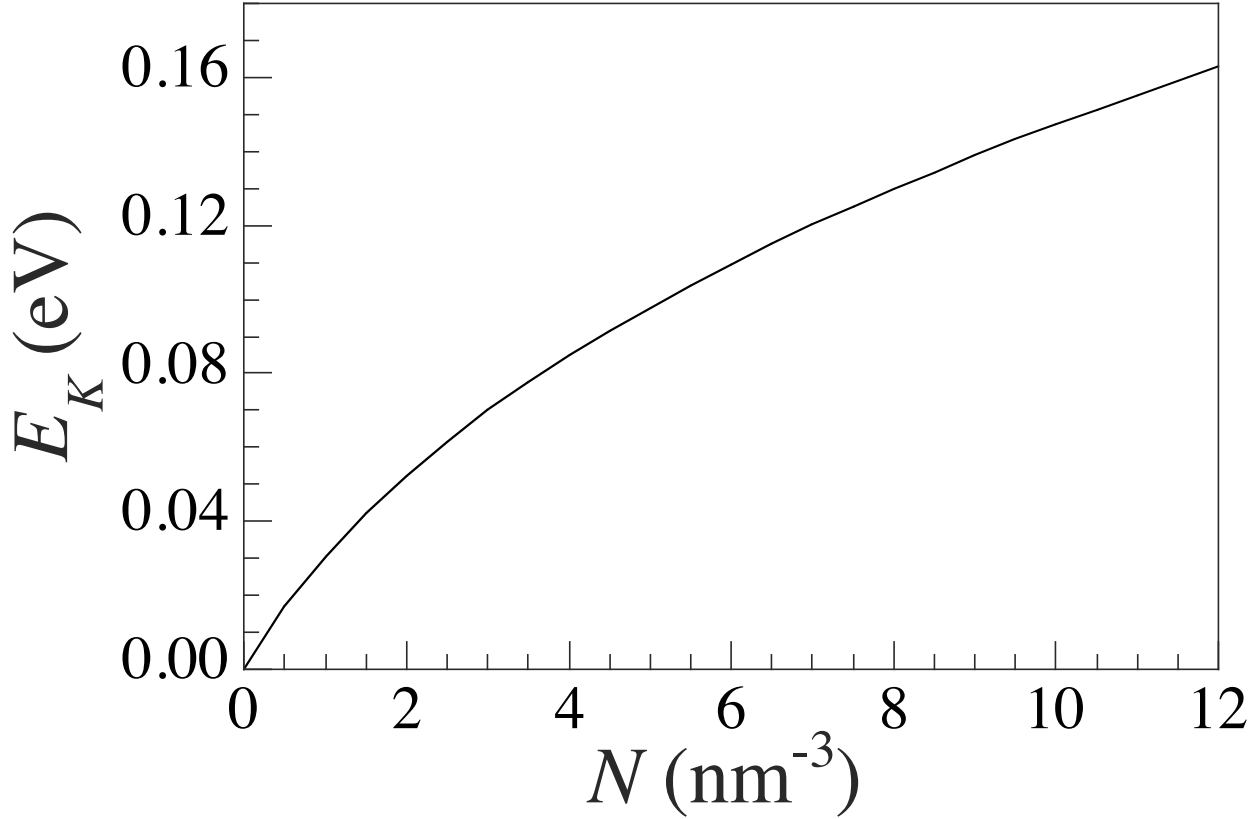}
\caption{\label{fig:ekdienne}\small $E_K(N)$ vs $N$ .}
\end{figure}

 As shown elsewhere, the electron-atom scattering in a dense gas is strongly affected by only $E_K$~\cite{bsl1992} that can be calculated by enforcing a local, average translational simmetry of the electron ground state wavefunction about the equivalent Wigner-Seitz (WS) cell~\cite{WS} centered on each gas atom~\cite{hernandezmartin1991}. This boundary condition yields the eigenvalue equation
 \begin{equation}
\tan{\left[k_0 \left(r_s-\tilde a\left(k_0\right)\right)\right]-k_0 r_s=0}
\label{eq:ee}
\end{equation}
that must be self-consistently solved for the wavevector $k_0 (N).$ $r_s=(3/4\pi N)^{1/3}$ is the radius of the WS cell. $\tilde a = (\sigma_t/4\pi)^{1/2}$ is the hard-core radius of the Hartree-Fock potential for rare gas atoms~\cite{SJC}, which is related to the total scattering cross section $\sigma_t.$
Once $k_0(N)$ has been determined, $E_K(N)$ is computed as
\begin{equation}
E_K(N)= \frac{\hbar^2}{2 m}k_0^2
\label{eq:ekn}
\end{equation}
In Fig.~\ref{fig:ekdienne} we show the dependence of $E_K$ on $N$ computed by means of Eq.~\ref{eq:ee} and Eq.~\ref{eq:ekn} by using the total scattering cross section given in literature~\cite{wey1988}.

\begin{figure}[t!]
\centering
\includegraphics[width=\columnwidth]{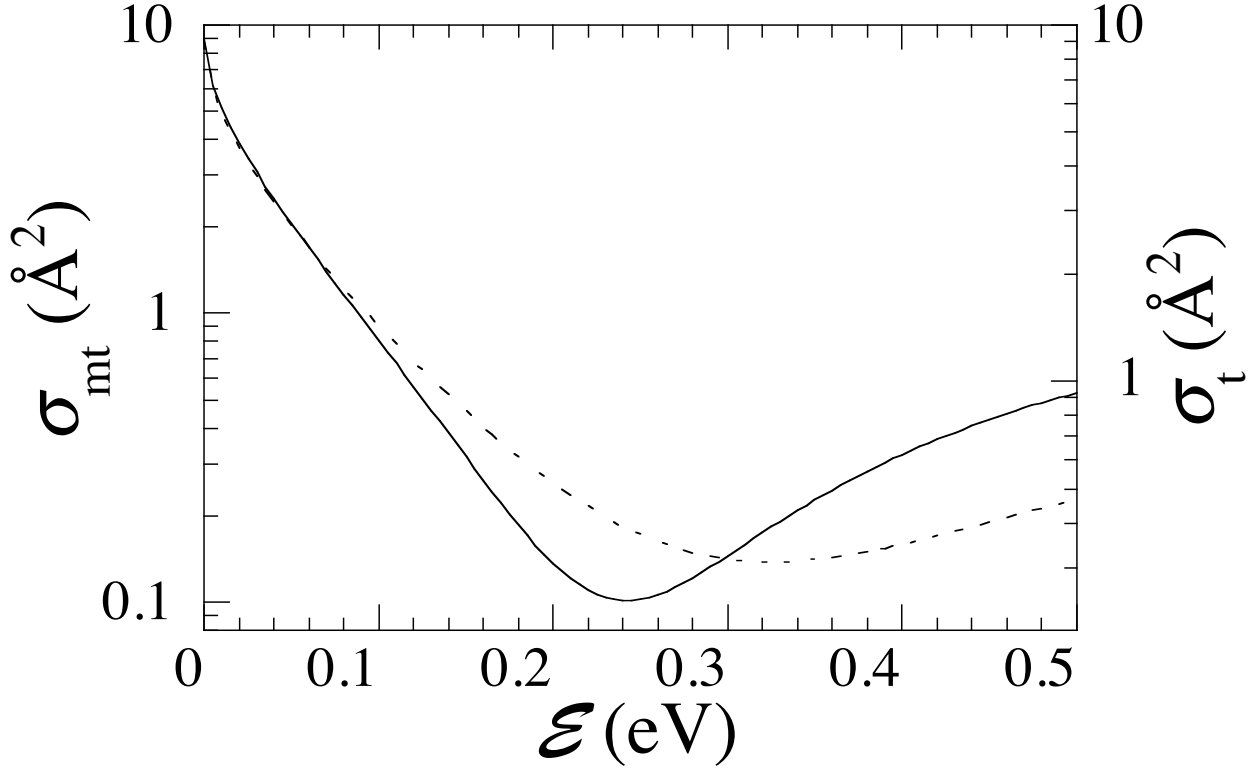}
\caption{\label{fig:CrossSectAr}\small Energy dependence of the electron-Ar scattering cross sections~\cite{wey1988}. Solid line: momentum transfer cross section (left scale). Dashed line: total cross section (right scale).}
\end{figure}

The experiments on electron mobility in dense noble gases~\cite{bsl1992,Bartels1973,borg1988,borg1990,lamp1990,Borghesani2001,lamp1994} have produced strong  evidence that the kinetic energy $\mathcal{E}$ of electrons during collisions is shifted by the amount $E_K(N),$ thereby producing a density dependence of the effective cross section through the energy dependence of the zero-density, electron-atom cross section itself that is shown in Fig.~\ref{fig:CrossSectAr}

\noindent It can be qualitatively understood, just by inspecting Fig.~\ref{fig:CrossSectAr}, that the effective cross section decreases if the average energy (namely, the thermal energy) is increased by $E_K (N)>0$ as $N$ is increased. For instance, at high density the energy shift is quite large, as shown in Fig.~\ref{fig:ekdienne}. So, the effective cross section may actually be much smaller than what would be computed at thermal energy. 

This is the main reason why the old data of Smejtek {\em et al.}~\cite{smejtek1973} so strongly disagree with the present results. In their analysis, they did not take into account the density dependent energy shift because its importance was not yet acknowledged and because a coherent physical picture of the electron scattering in dense gases was not yet emerged at that time. However, their results can be reconciled with the present ones if their raw data are now analyzed taking into account $E_K(N)$~\cite{borg2011}. 

For the sake of completeness, we have to mention that there are two more multiple scattering effects affecting the electron mobility in dense gases when electron mfp and thermal wavelength become comparable to each other. The first one is a quantum self-interference of the electron wave function scattered off atoms located along paths that are connected by time-reversal symmetry~\cite{ascarelli1986}, which enhances the backscattering rate~\cite{atrazhev1977,polischuk}. 

The second MS effect is an enhancement of the scattering cross section due to the strong correlations among scatterers, which set in close to the critical point. Actually, at fairly low $T$ and high $N,$ the electron wave function spreads over a wide region encompassing many atoms. The total scattered wave packet is obtained by coherently summing up all the partial scattering amplitudes contributed by each individual atom. As a result, the scattering cross section is enhanced by the static structure factor of the gas~\cite{lekner1967,cohen1967,lekner1968,lekner1973,steinberger1986} $S(0)=Nk_\mathrm{B}T \chi_T,$ where $\chi_T $ is the isothermal gas compressibility.

The latter two effects strongly affect the propagation of the wave packet and their influence is best observed in the way the electron mobility is modified with respect to the prediction of kinetic theory.  However, they are less effective in altering the electron energy distribution function. Thus, they can be neglected because the YB model assumes that electrons may be lost at their first encounter well before they can substantially propagate.

The electron energy distribution function $g(\mathcal{E})$ is given by the Davydov-Pidduck distribution~\cite{cohen1967,wannier}
\begin{eqnarray}
  g(\mathcal{E})&=& 
 C\exp\left\{
  -\int\limits_0^\mathcal{E}
  \Biggl[
  k_\mathrm{B}T   +\Bigr.
  \right.
 \nonumber \\
 &&\left. \left.+\frac{M}{6mz}
 \left(
 \frac{eE}{N\sigma_\mathrm{mt}\left(z+E_K (N)\right)}
 \right)^2\right]^{-1}\,\mathrm{d}z\right\}
\label{eq:gdie}
\end{eqnarray}
$k_\mathrm{B}$ is the Boltzmann's constant. $M$ and $m$ are the argon and electron masses, respectively. The constant $C$ is computed by enforcing the normalization condition $\int_0^\infty z^{1/2}g(z)\,\mathrm{d}z=1.$
In Eq.~\ref{eq:gdie} 
the dependence of the cross section on the shifted energy is explicitly shown. 

\begin{figure}\centering
\includegraphics[width=\columnwidth]{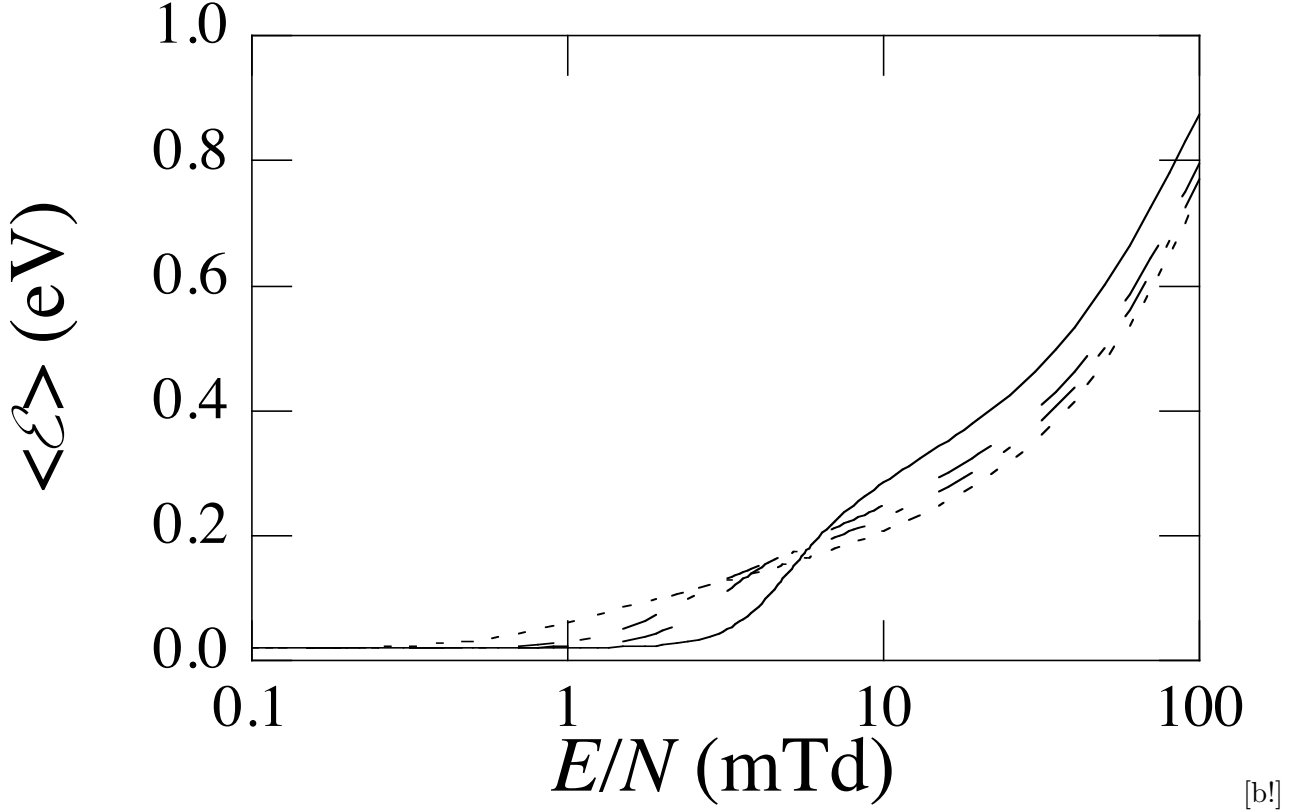}[b!]
\caption{\label{fig:MeanEnergy0e25e5e10}\small Reduced field dependence of the electron mean energy $\langle \mathcal{E}\rangle $ for some densities. 
$N (\mathrm{nm}^{-3})=0$ (solid line), $ 0.5$ (dashed line), $5$ (dash-dotted line) and 10 (dotted line). }
 \end{figure}

The knowledge of the distribution function allows one to compute averages. In Fig~\ref{fig:MeanEnergy0e25e5e10} we show how the mean electron energy $\langle \mathcal{E}\rangle$ depends on the reduced field $E/N$ for several densities.  
Except for the highest densities, we realize that the electron mean energy does not rise appreciably above its thermal value (except for the $E_K$ contribution) until quite high field strengths $E/N\approx 2\div 3\,$mTd are reached. This observation further suggests that the thermalization processes in argon are very fast.  

\begin{figure} 
\centering
\includegraphics[width=\columnwidth]{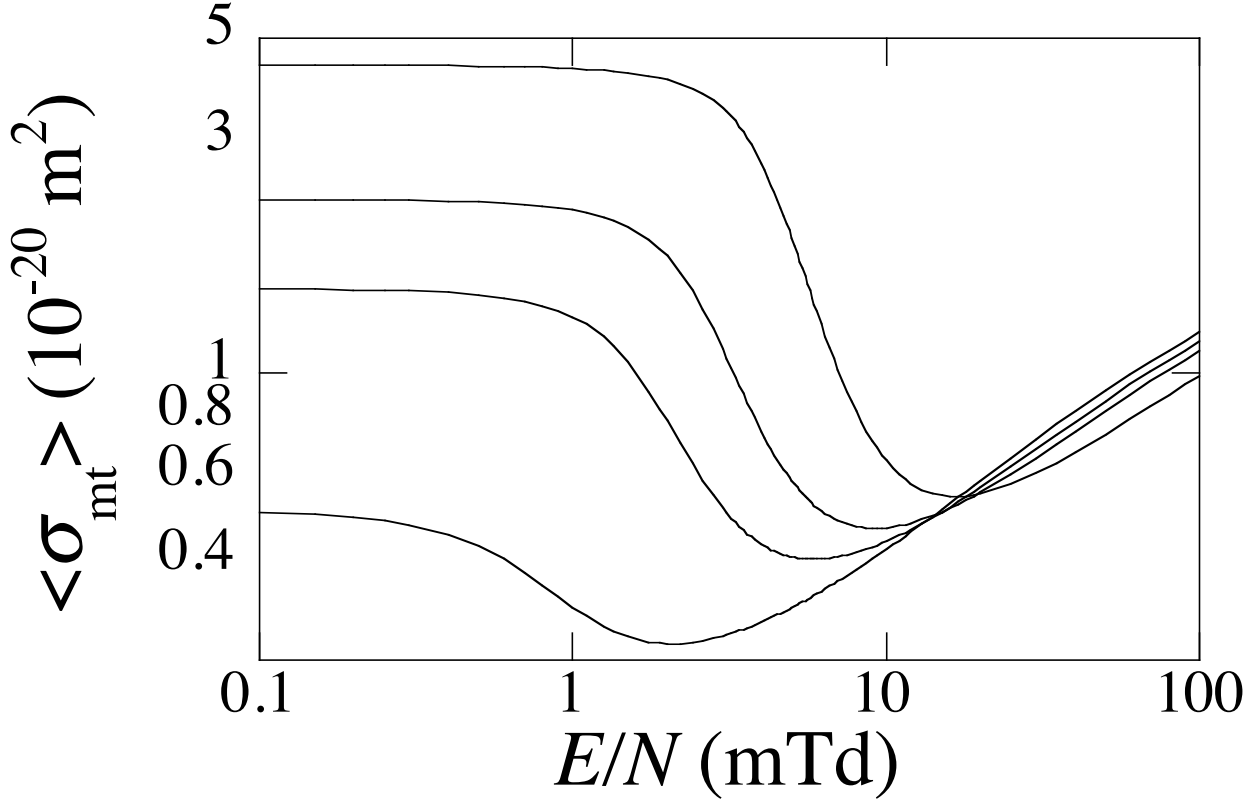}
\caption{\label{fig:sigmaave0e25e5e10}\small Reduced field dependence of the average momentum transfer scattering cross section $\langle \sigma_\mathrm{mt}\rangle $ evaluated for $T=152.7\,$K including the kinetic energy shift $E_{K.}$  $N (\mathrm{nm}^{-3})=0,$ $ 0.5,$ $5,$ and 10 (from top). }
\end{figure}

In Fig.~\ref{fig:sigmaave0e25e5e10} the average values of the momentum transfer cross section $\langle \sigma_\mathrm{mt}\rangle$ are plotted as a function of $E/N$ for several densities. 
The average cross section now exhibits a strong density dependence. Up to an intermediate field strength that depends on $N,$ $\langle \sigma_\mathrm{mt}\rangle $ turns out to be field independent and practically equal to its value at the shifted mean energy 
$\mathcal{\bar E}= \langle \mathcal{E}\rangle +E_K(N),$ i.e. $\langle \sigma_\mathrm{mt}\rangle \approx \sigma_\mathrm{mt}\left(\mathcal{\bar E}\right).$
  
The MS effects lead to a strong density dependence of the scattering cross section, especially at low fields. At high fields, this acquired density dependence diminishes. This fact is quite obvious because the electron wavelength becomes shorter as the field, hence the electron energy, increases, thus leading to a reduction of MS. 

The density dependence introduced by $E_{K}(N)$ into the cross section, combined with the strong energy dependence of  $\sigma_{\mathrm{mt}},$ leads to a strong reduction of the effective cross section with respect to the zero-density value. For instance, for $T=152.7\,$K and $N=0.5\,$nm$^{-3},$ $\langle \sigma_{\mathrm{mt}} \rangle\approx \sigma_{\mathrm{mt}}
(\mathcal{\bar E})\approx 2.3\times 10^{-20}\,$m$^{-2},$ whereas~\cite{wey1988} $\sigma_{\mathrm{mt}}((3/2)k_{\mathrm{B}}T)\approx 4.2\times 10^{-20}\,$m$^{-2}.$

In order to compare the results for the average, density dependent cross section predicted by the MS model with the density dependent value of the cross section determined from the collected charge data within the YB picture, we have calculated the effective cross section for $E/N\approx 1.5\,$mTd to take into account the fact that the charge data have been fitted to the YB model for $E/N\gtrsim 1 \,$mTd. 
This is a reasonable value corresponding  for almost all densities to the transition between thermal- and epithermal electron behavior. 

The cross section values calculated according to the MS model are plotted as solid lines in Fig.~\ref{fig:SigmaQT153and200}, in which they are compared with the experimental outcome. 
As reference densities for normalization purposes (see Eq.~\ref{eq:sigmanorm}) we have chosen $N_{0}=0.796\,$nm$^{-3}$ with $\sigma_{0}= 2.33\times 10^{-20}\,$m$^{2}$ for $T=200\, $K and $N_{0}=5.36\,$nm$^{-3}$ with $\sigma_{0}=1.01\times 10^{-20}\,$m$^{2}.$ The agreement between data and model is excellent up to the highest $N\approx 11.4\,$nm$^{-3}$ at the lower temperature. 

It is now clear that the charge $Q$ collected at fixed and moderately high $E/N$ in argon increases with $N$ because the momentum transfer cross section strongly decreases with increasing energy (see Fig.~\ref{fig:CrossSectAr}) leading to a decrease of the effective cross section with increasing $N.$
In neon, whose momentum transfer cross section rapidly increases with energy and whose effective cross section thus increases with $N$~\cite{borg1988}, $Q$ should decrease with increasing $N.$ This behavior has actually been observed~\cite{johnson1979}.

The present results confirm what has been obtained at different temperatures, both below~\cite{borg2011} and above~\cite{BLIEEE2012} $T_{c}.$ The disagremeent of the oldest data by Smejtek  {\em et al.}~\cite{smejtek1973} with the present results has been thouroughly discussed elsewhere~\cite{borg2011}. It suffices here to say that the old data are reconciled with the present results if they are more properly normalized to the most recent and accurate  mobility data~\cite{bsl1992}.

\subsubsection{Some final comments on the YB model}\label{sect:lowNdata}
The data presented here and elsewhere~\cite{borg2011,BLIEEE2012} span a wide range of reduced electric field data much lower than that explored with a different electron injection technique~\cite{smejtek1973}. The experimental results for (relatively) large field strengths, and also the old data at much higher field strengths (if more accurately normalized), agree well with the YB model, which predicts that a fraction of the epithermal electrons injected into the gas returns to the cathode immediately after being scattered off a gas atom in their first encounter. 

The cross section determined from the collected charge data within the YB model excellently agrees with the prediction of the MS model~\cite{bsl1992} that has been developed to treat the MS effects affecting the electron mobility in a dense noble gas. This good agreement lends credibility to the YB model whereas the MS model has been validated by its ability to describe the electron mobility data. 

However, it is not yet clear if the YB model is correct at all. Severe criticism has been raised against it. In particular, some of the fundamental hypotheses on which the YB model relies are strongly criticized. For instance,  the YB predictions are based on the assumption that electrons are backscattered to the cathode upon their first encounter. However, the prediction of the MS model is computed by taking thermal averages of the gas-phase cross section and by introducing MS effects. This mean that in the MS model it is implicitly assumed that electrons have reached thermal equilibrium with the gas. However, it is well known that electron thermalization occurs after a huge number of collisions~\cite{mozumder1980,kuntz1982}, thus overtly contradicting the most important hypothesis of the YB model.

MC calculations~\cite{kuntz1982,allen1984} confirm that a very large number of collisions are required to determine the fate of an electron, once again contradicting the YB hypothesis. However, they reproduce the $(E/N)^{1/2}$ behavior of the collected charge as a consequence of purely statistical effects~\cite{kuntz1982} but do not give any physical explanation of such a behavior. In addition to that, MC calculations fail at reproducing the density ordering of the experimental data because they are carried out for classical electron trajectories without taking into account the quantum MS effects which are at work in a dense gaseous environment.

\subsection{The low-field data}\label{sect:lowE}
In this section we present and discuss the data of collected charge at low field and at $T=152.7\,$K, very close to the critical temperature, $T_{c}\approx 150.9\,$K. 

Whereas the high-field data have been interpreted within a heuristic, MS model based on kinetic theory, we will show that the weak-field data show that a different physical phenomenon occurs. 

As previously noted~\cite{borg2011,BLIEEE2012}, the low-field data  show a density ordering that is opposite to the order at high fields. Actually, at high $E/N,$ the charge collected at fixed $E/N$ increases with increasing $N.$ If $E/N$ is lowered, there is a crossover region through which the ordering is reversed. In the previous experiments, we did not carry out measurements at such large densities to thoroughly explore this reversed density ordering. In one case, the temperature was below $T_{c}$ and we were limited by the density at coexistence in order to remain in the one-phase region. In the other case, $T\gg T_{c},$ and we were limited by the pressure the cell can withstand.

In the present experiment, we have been able to carry out measurements quite close to the critical temperature, thereby reaching densities comparable to or even larger than the density of the liquid.

In order to discuss these data, we refer to Fig.~\ref{fig:QNT15302and5mTd}. Whereas the charge collected at fixed $E/N=5\,$mTd, $Q_{5},$ steadily increases with $N,$ the charge collected for $E/N=0.2\, $mTd, $Q_{0.2},$ is not only smaller than $Q_{5}$ but decreases at first with increasing $N,$ shows a minimum about $N_{m}=6\,$nm$^{-3},$ and then it rises as $N$ is further increases.

\subsubsection{Electrostriction}\label{sect:ES}
\begin{figure}[t!]
\includegraphics[width=\columnwidth]{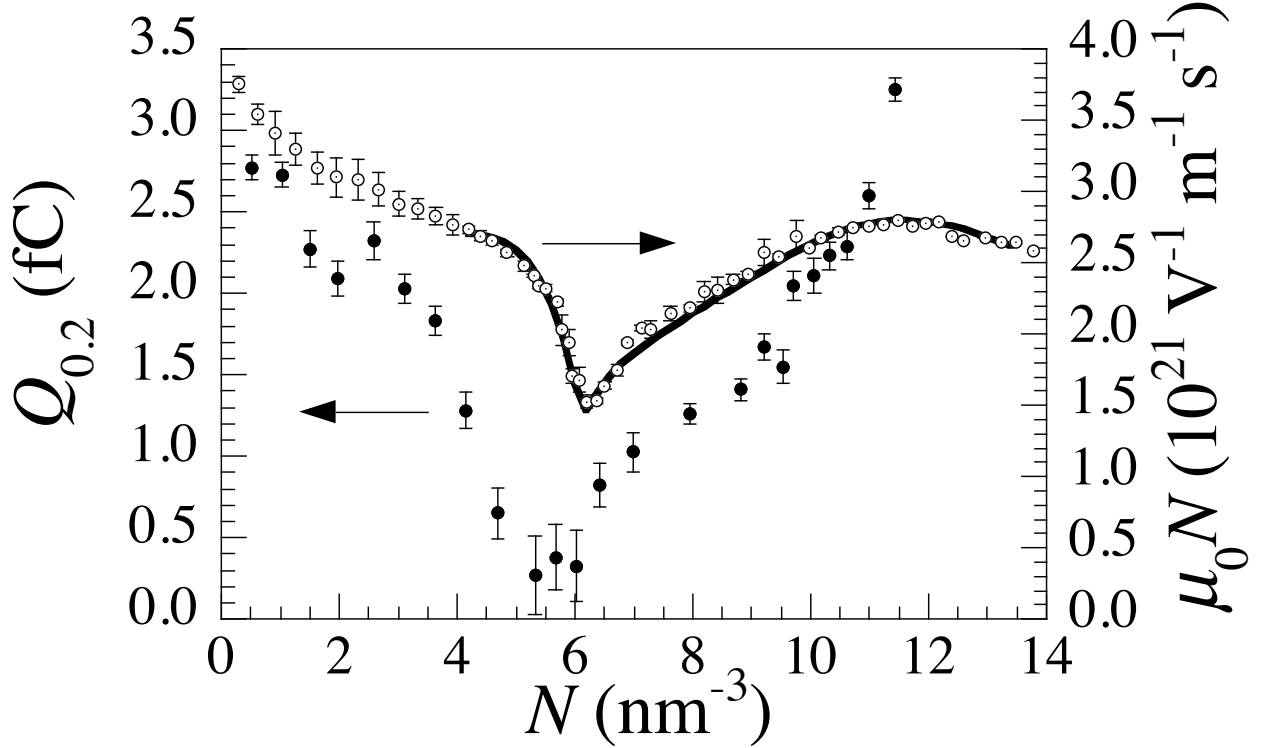}
\caption{\label{fig:QEN02T153MIU0N151} \small Density dependence of the charge extracted at weak fields $Q_{0.2} $ (closed circles, left scale) and of the zero-field density-normalized mobility $\mu_{0}N$ of O$_{2}^{-}$ ions~\cite{cpl} (open circles, right scale). Solid line: prediction of the hydrodynamic Stokes's formula for the ion mobility~\cite{byron} using an effective ion radius that includes contributions from solvation and critical point correlation effects.
 Critical density: $N_{c}\approx 8.08\,$nm$^{-3}.$ }
\end{figure}

The behavior of $Q_{0.2}$ shown in Fig.~\ref{fig:QNT15302and5mTd} closely resembles the behavior of the zero-field density-normalized mobility $\mu_{0}N$ of O$_{2}^{-} $ ions in argon gas for $T=151.5\,$K~\cite{cpl,o2m}, a temperature very close to the present one and even closer to $T_{c}.$ In Fig.~\ref{fig:QEN02T153MIU0N151}
we plot $Q_{0.2}$ (left scale) and $\mu_{0} N$ (right scale) as a function of $N.$
Both $Q_{0.2}$ and $\mu_{0}N$ have a minimum at practically the same density, $N_{m}\approx 6\,$nm$^{3}<N_{c}=8.08\,$nm$^{-3}.$

The behavior of the ion mobility is explained in terms of electrostriction~\cite{atkins1959,neri1993}. 
The ion electric field polarizes the surrounding gas atoms. As ions are drifting very slowly through the dense gas with a typical drift velocity of a few mm/s even at moderately high field strength~\cite{cpl}, the gas has time to react to the external perturbation by enhancing the local density around the ion. The solvation shell that is so built around the ion has to be dragged with it during its motion, thereby increasing its hydrodynamic resistance and reducing its mobility. 

Owing to the far larger drift velocity of electrons with respect to that of ions, one would expect electrostriction not to play a major r\^ole in the electron behavior. This is normally true. However, in our experiment, at weak fields, electrons may drift with a velocity smaller than the sound velocity that determines the gas response to perturbations. 

\begin{figure}[b!]
\centering
\includegraphics[width=\columnwidth]{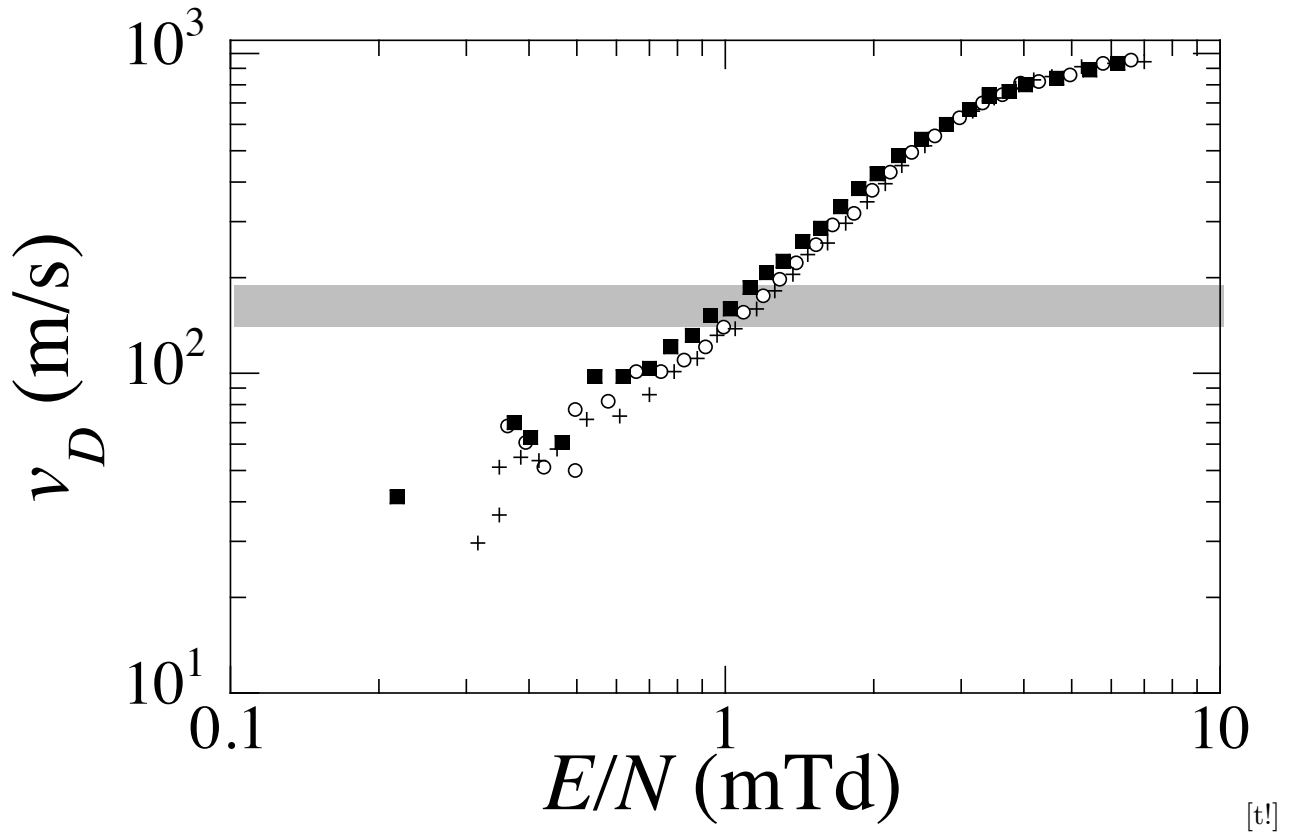}[t!]
\caption{\label{fig:VdLMN153}\small $v_{D}$ vs $E/N$ for $N$ close to the density at which the collected charge at low field is minimum. $N\,(\mbox{nm}^{-3})=$ 6.43 (squares), 6.02 (circles), and 5.67 (crosses)~\cite{Borghesani2001}. The hatched region roughly corresponds to the range of sound speed in the near critical region~\cite{thoen1971}.}
\end{figure}

In Fig.~\ref{fig:VdLMN153} the electron drift velocity $v_{D}$~\cite{Borghesani2001} is plotted as a function of the reduced field $E/N$ for the densities close to $N_{m},$ where $Q_{0.2}$ shows the minimum.
The grey band in the figure represents the range of sound speed in the corresponding density and temperature region~\cite{thoen1971}.

It can immediately be realized that the electron drift velocity at weak field, in particular for $E/N\approx 0.2\,$mTd, is much smaller than the sound velocity. The obvious consequence is that the medium can react also to the perturbation induced by the electric field of the electrons. The medium response, thus, leads to the formation of a solvation shell surrounding the electrons. Once electrons are solvated, their motion is obviously hindered and they are less prone to be collected at the anode. 
This is the physical reason of the initial decrease of the collected charge with increasing density.

It is also interesting to note that no such solvation effects are observed in the drift mobility~\cite{lamp1990,Borghesani2001}. Actually, the mobility is measured by detecting only those electrons that are not solvated and can propagate across the whole drift distance, thereby reaching the anode. On the other hand, solvated electrons are returned to the cathode and do not contribute to the observed mobility. 

For densities $N\approx N_{m}<N_{c},$ $Q_{0.2}$ shows a minimum and, then, it increases again for $N>N_{m}.$ If the model of the (partial) formation of a solvation shell surrounding a fraction of the photoinjected electrons is accepted, one would na\"{\i}vely expect that the strongest effect should occur at the critical density, at which the gas compressibility is the largest. On the contrary, the experimental evidence is that the maximum electrostriction effect occurs for $N_{m}<N_{c}.$ 

In order to explain this observation and the further rise of $Q_{0.2}$ with increasing density for $N>N_{m},$ 
it is worth working out into some numerical details the electrostriction model developed by Atkins~\cite{atkins1959}.
\subsubsection{Equations and results of the electrostriction model}\label{sect:esmeq}
The gas is treated as a classical continuum described by a density dependent relative dielectric constant $K(N)$ that is related to the atomic polarizability $\alpha\>(\approx 1.8\times 10^{-40}\,$F$\cdot$m for Ar~\cite{maitland}) by the well-known Clausius-Mossotti relationship~\cite{kittel}
\begin{equation}
\frac{K-1}{K+2}= \frac{N\alpha}{3\epsilon_{0}}
\label{eq:CM}\end{equation}
If the system is in thermal equilibrium, the chemical potential must be uniform everywhere. 
Let $g_{0}(T,P)$ be the chemical potential of the gas atoms in the unperturbed region very far away from the charge, where the gas density is $N_{\infty}.$
At a distance $r$ from the charge the electric field generated by it is $\mathbf{E}_{i}(r)$ and the local value of the chemical potential $g(r)$ must satisfy the condition
\begin{equation}
g(r)= g_{0} -\frac{1}{2}\frac{\mathbf{P}\cdot \mathbf{E}_{i}}{N}
\label{eq:chempot}\end{equation}
where $\mathbf{P}$ is the polarization of the medium.
As $g$ and $T$ must be constant throughout the gas in equilibrium~\cite{guggenheim}, differentiation of Eq.~\ref{eq:chempot} yields 
\begin{equation}
\frac{\mathrm{d}p}{N} =\frac{1}{N} \left( \frac{\partial p}{\partial N}\right)_{T}\,\mathrm{d}N= \frac{\mathbf{P}\cdot \mathrm{d}\mathbf{E}_{i}}{N}
\label{eq:dg0}\end{equation}
in which $p$ is the gas pressure.
Integration of Eq.~\ref{eq:dg0} gives
\begin{equation}
\int\limits_{N_{\infty}}^{N(r)} \frac{1}{N} \left( \frac{\partial p}{\partial N}\right)_{T}\,\mathrm{d}N= \frac{1}{2} \alpha E_{i}^{2}(r)
\label{eq:dgint}\end{equation}
where we have taken into account that the screening factor necessary to evaluate the field at the atom $3/[2+K(N)]\approx 1$ at all densities.

If the actual expression for $E_{i}= -e/[4\pi\epsilon_{0} K(N) r^{2}]$ is substituted into Eq.~\ref{eq:dgint}, we get
\begin{equation}
K^{2}(N)\int\limits_{N_{\infty}}^{N} \frac{1}{N} \left( \frac{\partial p}{\partial N}\right)_{T}\,\mathrm{d}N =
\frac{1}{2}\frac{\alpha e^{2}}{\left(4\pi\epsilon_{0}\right)^{2}r^{4}}
\label{eq:Gcal}\end{equation}
The l.h.s. of Eq.~\ref{eq:Gcal} is a function of only $N$ at constant $T$ and can be numerically evaluated if the equation of state of the gas is known. 
Eq.~\ref{eq:Gcal} can then be inverted so as to implicitly yield the density profile $N(r)$ around the charge. 
\begin{figure}[b!]
\centering
\includegraphics[width=\columnwidth]{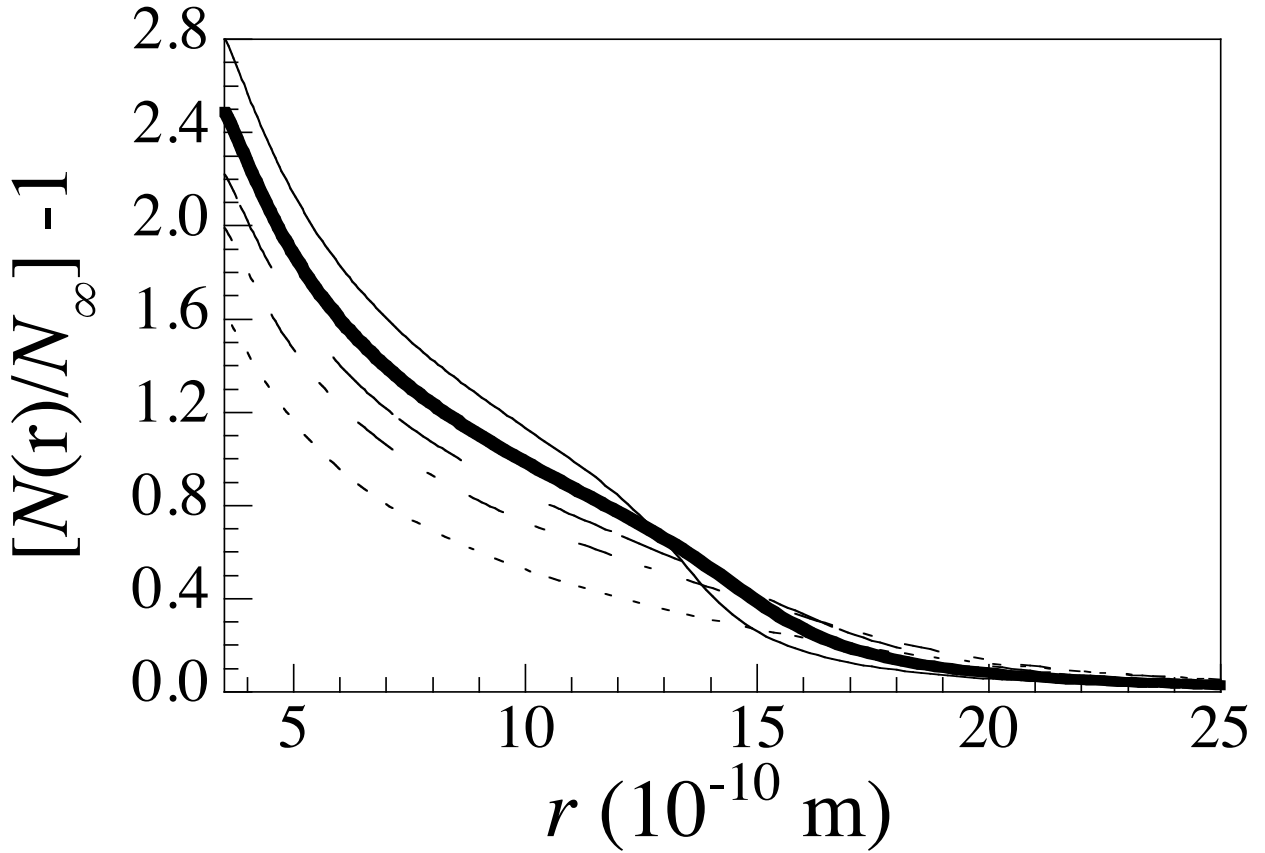}
\caption{\label{fig:profili153K}\small Pair correlation function $h(r)=[N(r)/N_{\infty}]-1$ vs $r$ for several unperturbed gas densities $N_{\infty}\,(\mathrm{nm}^{-3})=5.5$ (thin solid line), $6$ (thick solid line), $6.5$ (dashed line), $7$ (dash-dotted line) and $8$ (dotted line). The curves are computed for $T=152.7\,$K.  }
\end{figure}

In Fig.~\ref{fig:profili153K} we show the pair correlation function~\cite{chandler} $h(r) =[N(r)/N_{\infty}]-1$ for some densities of the unperturbed fluid. 
If the density of the unperturbed fluid is below the critical value $N_{c}, $ the local density $N(r)$ meets the value $N_{c}$ at some distance from the charge. There, the local compressibility is the largest.

On the contrary, if the unperturbed gas density is already above $N_{c},$ the increase of the local density as $r\rightarrow 0$ brings the gas to a condition in which the local compressibility is by far smaller than for $N_{c}.$ 

The pair correlation functions $h(r)$ calculated for some $N_{\infty}\equiv N $ around $N_{c}$ and plotted in Fig.~\ref{fig:profili153K} further show that for $N\approx 6\,$nm$^{-3}$ the inflexion point, i.e., the point at which the compressibility is the largest, occurs at a distance from the charge larger than for all other densities.

For instance, for $N=5.5\,$nm$^{-3}$ the slope, hence the compressibility, at the inflexion point is larger than for $N=6\,$nm$^{-3}\approx N_{m}$ but the inflexion point itself appears closer to the charge than for $N=6\,$nm$^{-3}.$ This observation leads to the conclusion that for $N=N_{m}\approx 6\,$nm$^{-3}$ the largest amount of gas is affected by the perturbation induced by the presence of the electron charge. 
For $N>N_{m}$ the inflexion rapidly disappears leading to a decrease of the gas response. 

Now, the picture is clear. When the density increases from below towards $N_{m},$ the perturbation induced by the charge involves a growing volume of gas. The maximum effect, a combination of strong gas compressibility and wide extension of the perturbation, is reached for $N_{m}\approx 6\,$nm$^{-3}<N_{c}. $ For even larger $N,$ the compressibility keeps decreasing and the response of the gas is less and less effective, thus leading to the resumption of the growth of the collected charge.

The case of the ion mobility is a fortunate one because the Stokes's hydrodynamic formula~\cite{byron} can be used to relate the extension of the gas perturbation, that includes both solvation and critical correlation effects~\cite{cpl,o2m} and that can be estimated by using the electrostriction model, to the measured mobility. Actually, as shown by the solid line in Fig.~\ref{fig:QEN02T153MIU0N151}, the agreement of the electrostriction model with the experimental data is very good.
Unfortunately, there is not any model relating the dimension of the solvation shell to the amount of charge that is returned to the cathode or, equivalently, to that collected at the anode. Thus, we must content ourselves with such a semiquantitative physical explanation of the experimental outcome.

\section{\label{sec:Conc}Conclusions} 
The present measurements of the collection efficiency of electrons photoinjected into dense argon gas have extended the density range and temperature of our previous measurements. 

We have confirmed that the YB model, in spite of its flaws, allows an independent determination of the effective, density-dependent momentum transfer scattering cross section that agrees with its determination from electron mobility measurements. 
This effective cross section can suitably be computed from the 
zero density, electron-atom cross section 
by taking into account its energy dependence and the MS effects, namely the density-dependent quantum shift of the energy at the bottom of the conduction band, the correlations among scatterers, and the backscattering rate enhancement due to quantum self-interference of the electron wave packet.

At the same time, the measurements close to the critical temperature have allowed us to reach very large density values. We have observed that, at low electric fields, the density dependence of the collected charge at fixed reduced field strength is different than that occurring at high field strength. We have interpreted this phenomenon as due to the formation of a solvation (or, correlation) shell of gas atoms around the electrons because they drift with a velocity much smaller than the sound velocity. A similar situation occurs for slow O$_{2}^{-}$ ions in near critical argon gas.

Unfortunately, we can only give a qualitative explanation of the experimental observations in this regime because we are not aware of a theoretical model that might extend the YB model in a thermodynamic region, in which the gas cannot be considered to be a uniform medium.

We believe that this kind of measurements together with the theoretical explanation we are proposing should be verified by measuring the electron collection efficiency also in other noble gases. 
Such measurements have been carried out in liquid helium but the outcome is determined by the fact that helium presents a $\approx 1\,$eV high barrier to electron injection~\cite{silver1967,onn1969,broomall1976}. 
Scanty experimental measurements have been carried out in neon gas and liquid~\cite{johnson1979}, which apparently confirm the explanation proposed in this paper. Unfortunately, the experimental accuracy is far to be satisfactory to get quantitatively significant results.


\begin{thebibliography}{72}%
\makeatletter
\providecommand \@ifxundefined [1]{%
 \@ifx{#1\undefined}
}%
\providecommand \@ifnum [1]{%
 \ifnum #1\expandafter \@firstoftwo
 \else \expandafter \@secondoftwo
 \fi
}%
\providecommand \@ifx [1]{%
 \ifx #1\expandafter \@firstoftwo
 \else \expandafter \@secondoftwo
 \fi
}%
\providecommand \natexlab [1]{#1}%
\providecommand \enquote  [1]{``#1''}%
\providecommand \bibnamefont  [1]{#1}%
\providecommand \bibfnamefont [1]{#1}%
\providecommand \citenamefont [1]{#1}%
\providecommand \href@noop [0]{\@secondoftwo}%
\providecommand \href [0]{\begingroup \@sanitize@url \@href}%
\providecommand \@href[1]{\@@startlink{#1}\@@href}%
\providecommand \@@href[1]{\endgroup#1\@@endlink}%
\providecommand \@sanitize@url [0]{\catcode `\\12\catcode `\$12\catcode
  `\&12\catcode `\#12\catcode `\^12\catcode `\_12\catcode `\%12\relax}%
\providecommand \@@startlink[1]{}%
\providecommand \@@endlink[0]{}%
\providecommand \url  [0]{\begingroup\@sanitize@url \@url }%
\providecommand \@url [1]{\endgroup\@href {#1}{\urlprefix }}%
\providecommand \urlprefix  [0]{URL }%
\providecommand \Eprint [0]{\href }%
\providecommand \doibase [0]{http://dx.doi.org/}%
\providecommand \selectlanguage [0]{\@gobble}%
\providecommand \bibinfo  [0]{\@secondoftwo}%
\providecommand \bibfield  [0]{\@secondoftwo}%
\providecommand \translation [1]{[#1]}%
\providecommand \BibitemOpen [0]{}%
\providecommand \bibitemStop [0]{}%
\providecommand \bibitemNoStop [0]{.\EOS\space}%
\providecommand \EOS [0]{\spacefactor3000\relax}%
\providecommand \BibitemShut  [1]{\csname bibitem#1\endcsname}%
\let\auto@bib@innerbib\@empty
\bibitem [{\citenamefont {Schmidt}(1997)}]{schmidtCRC}%
  \BibitemOpen
  \bibfield  {author} {\bibinfo {author} {\bibfnamefont {W.~F.}\ \bibnamefont
  {Schmidt}},\ }\href@noop {} {\emph {\bibinfo {title} {Liquid state
  electronics of insulating liquids}}}\ (\bibinfo  {publisher} {CRC Press},\
  \bibinfo {address} {Boca Raton},\ \bibinfo {year} {1997})\BibitemShut
  {NoStop}%
\bibitem [{\citenamefont {Tauchert}, \citenamefont {Jungblut},\ and\
  \citenamefont {Schmidt}(1977)}]{schmidt1977}%
  \BibitemOpen
  \bibfield  {author} {\bibinfo {author} {\bibfnamefont {W.}~\bibnamefont
  {Tauchert}}, \bibinfo {author} {\bibfnamefont {H.}~\bibnamefont {Jungblut}},
  \ and\ \bibinfo {author} {\bibfnamefont {W.~F.}\ \bibnamefont {Schmidt}},\
  }\href@noop {} {\bibfield  {journal} {\bibinfo  {journal} {Can. J. Chem.}\
  }\textbf {\bibinfo {volume} {55}},\ \bibinfo {pages} {1860} (\bibinfo {year}
  {1977})}\BibitemShut {NoStop}%
\bibitem [{\citenamefont {Borghesani}\ \emph {et~al.}(1986)\citenamefont
  {Borghesani}, \citenamefont {Bruschi}, \citenamefont {Santini},\ and\
  \citenamefont {Torzo}}]{borg1986}%
  \BibitemOpen
  \bibfield  {author} {\bibinfo {author} {\bibfnamefont {A.~F.}\ \bibnamefont
  {Borghesani}}, \bibinfo {author} {\bibfnamefont {L.}~\bibnamefont {Bruschi}},
  \bibinfo {author} {\bibfnamefont {M.}~\bibnamefont {Santini}}, \ and\
  \bibinfo {author} {\bibfnamefont {G.}~\bibnamefont {Torzo}},\ }\href
  {\doibase 10.1063/1.1138689} {\bibfield  {journal} {\bibinfo  {journal} {Rev.
  Sci. Instrum.}\ }\textbf {\bibinfo {volume} {57}},\ \bibinfo {pages} {2234}
  (\bibinfo {year} {1986})}\BibitemShut {NoStop}%
\bibitem [{\citenamefont {Silver}, \citenamefont {Smejtek},\ and\ \citenamefont
  {Masuda}(1967)}]{silver1967}%
  \BibitemOpen
  \bibfield  {author} {\bibinfo {author} {\bibfnamefont {M.}~\bibnamefont
  {Silver}}, \bibinfo {author} {\bibfnamefont {P.}~\bibnamefont {Smejtek}}, \
  and\ \bibinfo {author} {\bibfnamefont {K.}~\bibnamefont {Masuda}},\
  }\href@noop {} {\bibfield  {journal} {\bibinfo  {journal} {Phys. Rev. Lett.}\
  }\textbf {\bibinfo {volume} {19}},\ \bibinfo {pages} {626} (\bibinfo {year}
  {1967})}\BibitemShut {NoStop}%
\bibitem [{\citenamefont {Silver}\ \emph {et~al.}(1970)\citenamefont {Silver},
  \citenamefont {Kumbhare}, \citenamefont {Smjtek},\ and\ \citenamefont
  {Onn}}]{onn1970}%
  \BibitemOpen
  \bibfield  {author} {\bibinfo {author} {\bibfnamefont {M.}~\bibnamefont
  {Silver}}, \bibinfo {author} {\bibfnamefont {P.}~\bibnamefont {Kumbhare}},
  \bibinfo {author} {\bibfnamefont {P.}~\bibnamefont {Smjtek}}, \ and\ \bibinfo
  {author} {\bibfnamefont {D.~G.}\ \bibnamefont {Onn}},\ }\href@noop {}
  {\bibfield  {journal} {\bibinfo  {journal} {J. Chem. Phys.}\ }\textbf
  {\bibinfo {volume} {52}},\ \bibinfo {pages} {5195} (\bibinfo {year}
  {1970})}\BibitemShut {NoStop}%
\bibitem [{\citenamefont {Blossey}(1974)}]{Blossey1974}%
  \BibitemOpen
  \bibfield  {author} {\bibinfo {author} {\bibfnamefont {D.~F.}\ \bibnamefont
  {Blossey}},\ }\href {\doibase 10.1103/PhysRevB.9.5183} {\bibfield  {journal}
  {\bibinfo  {journal} {Phys. Rev. B}\ }\textbf {\bibinfo {volume} {9}},\
  \bibinfo {pages} {5183} (\bibinfo {year} {1974})}\BibitemShut {NoStop}%
\bibitem [{\citenamefont {Smejtek}\ \emph {et~al.}(1973)\citenamefont
  {Smejtek}, \citenamefont {Silver}, \citenamefont {Dy},\ and\ \citenamefont
  {Onn}}]{smejtek1973}%
  \BibitemOpen
  \bibfield  {author} {\bibinfo {author} {\bibfnamefont {P.}~\bibnamefont
  {Smejtek}}, \bibinfo {author} {\bibfnamefont {M.}~\bibnamefont {Silver}},
  \bibinfo {author} {\bibfnamefont {K.~S.}\ \bibnamefont {Dy}}, \ and\ \bibinfo
  {author} {\bibfnamefont {D.~G.}\ \bibnamefont {Onn}},\ }\href@noop {}
  {\bibfield  {journal} {\bibinfo  {journal} {J. Chem. Phys.}\ }\textbf
  {\bibinfo {volume} {59}},\ \bibinfo {pages} {1374} (\bibinfo {year}
  {1973})}\BibitemShut {NoStop}%
\bibitem [{\citenamefont {Huxley}\ and\ \citenamefont
  {Crompton}(1974)}]{Huxley}%
  \BibitemOpen
  \bibfield  {author} {\bibinfo {author} {\bibfnamefont {L.~G.~H.}\
  \bibnamefont {Huxley}}\ and\ \bibinfo {author} {\bibfnamefont {R.~W.}\
  \bibnamefont {Crompton}},\ }\href@noop {} {\emph {\bibinfo {title} {Diffusion
  and drift of electrons in gases}}}\ (\bibinfo  {publisher} {Wiley},\ \bibinfo
  {address} {New York},\ \bibinfo {year} {1974})\BibitemShut {NoStop}%
\bibitem [{\citenamefont {Young}\ and\ \citenamefont
  {Bradbury}(1933)}]{YB1933}%
  \BibitemOpen
  \bibfield  {author} {\bibinfo {author} {\bibfnamefont {L.~A.}\ \bibnamefont
  {Young}}\ and\ \bibinfo {author} {\bibfnamefont {N.~E.}\ \bibnamefont
  {Bradbury}},\ }\href {\doibase 10.1103/PhysRev.43.34} {\bibfield  {journal}
  {\bibinfo  {journal} {Phys. Rev.}\ }\textbf {\bibinfo {volume} {43}},\
  \bibinfo {pages} {34} (\bibinfo {year} {1933})}\BibitemShut {NoStop}%
\bibitem [{\citenamefont {Borghesani}\ and\ \citenamefont
  {Lamp}(2011)}]{borg2011}%
  \BibitemOpen
  \bibfield  {author} {\bibinfo {author} {\bibfnamefont {A.~F.}\ \bibnamefont
  {Borghesani}}\ and\ \bibinfo {author} {\bibfnamefont {P.}~\bibnamefont
  {Lamp}},\ }\href@noop {} {\bibfield  {journal} {\bibinfo  {journal} {Plasma
  Sources Sci. Technol}\ }\textbf {\bibinfo {volume} {20}},\ \bibinfo {pages}
  {034001} (\bibinfo {year} {2011})}\BibitemShut {NoStop}%
\bibitem [{\citenamefont {Borghesani}\ and\ \citenamefont
  {Lamp}(2012)}]{BLIEEE2012}%
  \BibitemOpen
  \bibfield  {author} {\bibinfo {author} {\bibfnamefont {A.~F.}\ \bibnamefont
  {Borghesani}}\ and\ \bibinfo {author} {\bibfnamefont {P.}~\bibnamefont
  {Lamp}},\ }\href {\doibase 10.1109/TDEI.2012.6180264} {\bibfield  {journal}
  {\bibinfo  {journal} {IEEE Trans. Dielect. Electr. Insul.}\ }\textbf
  {\bibinfo {volume} {19}},\ \bibinfo {pages} {689} (\bibinfo {year}
  {2012})}\BibitemShut {NoStop}%
  \bibitem [{\citenamefont {Borghesani}, \citenamefont {Santini},\ and\
  \citenamefont {Lamp}(1992)}]{bsl1992}%
  \BibitemOpen
  \bibfield  {author} {\bibinfo {author} {\bibfnamefont {A.~F.}\ \bibnamefont
  {Borghesani}}, \bibinfo {author} {\bibfnamefont {M.}~\bibnamefont {Santini}},
  \ and\ \bibinfo {author} {\bibfnamefont {P.}~\bibnamefont {Lamp}},\ }\href
  {\doibase 10.1103/PhysRevA.46.7902} {\bibfield  {journal} {\bibinfo
  {journal} {Phys. Rev. A}\ }\textbf {\bibinfo {volume} {46}},\ \bibinfo
  {pages} {7902} (\bibinfo {year} {1992})}\BibitemShut {NoStop}%
\bibitem [{\citenamefont {Kuntz}\ and\ \citenamefont
  {Schmidt}(1982)}]{kuntz1982}%
  \BibitemOpen
  \bibfield  {author} {\bibinfo {author} {\bibfnamefont {P.~J.}\ \bibnamefont
  {Kuntz}}\ and\ \bibinfo {author} {\bibfnamefont {W.~F.}\ \bibnamefont
  {Schmidt}},\ }\href {\doibase 10.1063/1.443082} {\bibfield  {journal}
  {\bibinfo  {journal} {J. Chem. Phys.}\ }\textbf {\bibinfo {volume} {76}},\
  \bibinfo {pages} {1136} (\bibinfo {year} {1982})}\BibitemShut {NoStop}%
\bibitem [{\citenamefont {Borghesani}\ \emph {et~al.}(1988)\citenamefont
  {Borghesani}, \citenamefont {Bruschi}, \citenamefont {Santini},\ and\
  \citenamefont {Torzo}}]{borg1988}%
  \BibitemOpen
  \bibfield  {author} {\bibinfo {author} {\bibfnamefont {A.~F.}\ \bibnamefont
  {Borghesani}}, \bibinfo {author} {\bibfnamefont {L.}~\bibnamefont {Bruschi}},
  \bibinfo {author} {\bibfnamefont {M.}~\bibnamefont {Santini}}, \ and\
  \bibinfo {author} {\bibfnamefont {G.}~\bibnamefont {Torzo}},\ }\href
  {\doibase 10.1103/PhysRevA.37.4828} {\bibfield  {journal} {\bibinfo
  {journal} {Phys. Rev. A}\ }\textbf {\bibinfo {volume} {37}},\ \bibinfo
  {pages} {4828} (\bibinfo {year} {1988})}\BibitemShut {NoStop}%
\bibitem [{\citenamefont {Borghesani}\ and\ \citenamefont
  {Santini}(1990{\natexlab{a}})}]{borg1990}%
  \BibitemOpen
  \bibfield  {author} {\bibinfo {author} {\bibfnamefont {A.~F.}\ \bibnamefont
  {Borghesani}}\ and\ \bibinfo {author} {\bibfnamefont {M.}~\bibnamefont
  {Santini}},\ }\href {\doibase 10.1103/PhysRevA.42.7377} {\bibfield  {journal}
  {\bibinfo  {journal} {Phys. Rev. A}\ }\textbf {\bibinfo {volume} {42}},\
  \bibinfo {pages} {7377} (\bibinfo {year} {1990}{\natexlab{a}})}\BibitemShut
  {NoStop}%
\bibitem [{\citenamefont {Borghesani}\ and\ \citenamefont
  {Santini}(2002)}]{borg2002}%
  \BibitemOpen
  \bibfield  {author} {\bibinfo {author} {\bibfnamefont {A.~F.}\ \bibnamefont
  {Borghesani}}\ and\ \bibinfo {author} {\bibfnamefont {M.}~\bibnamefont
  {Santini}},\ }\href {\doibase 10.1103/PhysRevE.65.056403} {\bibfield
  {journal} {\bibinfo  {journal} {Phys. Rev. E}\ }\textbf {\bibinfo {volume}
  {65}},\ \bibinfo {pages} {056403} (\bibinfo {year} {2002})}\BibitemShut
  {NoStop}%
\bibitem [{\citenamefont {Hernandez}(1991)}]{Hernandez1991}%
  \BibitemOpen
  \bibfield  {author} {\bibinfo {author} {\bibfnamefont {J.~P.}\ \bibnamefont
  {Hernandez}},\ }\href {\doibase 10.1103/RevModPhys.63.675} {\bibfield
  {journal} {\bibinfo  {journal} {Rev. Mod. Phys.}\ }\textbf {\bibinfo {volume}
  {63}},\ \bibinfo {pages} {675} (\bibinfo {year} {1991})}\BibitemShut
  {NoStop}%
\bibitem [{\citenamefont {Borghesani}(2001)}]{Borghesani2001}%
  \BibitemOpen
  \bibfield  {author} {\bibinfo {author} {\bibfnamefont {A.~F.}\ \bibnamefont
  {Borghesani}},\ }\href {\doibase 10.1016/S0304-3886(01)00133-4} {\bibfield
  {journal} {\bibinfo  {journal} {J. Electrostatics}\ }\textbf {\bibinfo
  {volume} {53}},\ \bibinfo {pages} {89} (\bibinfo {year} {2001})}\BibitemShut
  {NoStop}%
\bibitem [{\citenamefont {Lekner}(1967)}]{lekner1967}%
  \BibitemOpen
  \bibfield  {author} {\bibinfo {author} {\bibfnamefont {J.}~\bibnamefont
  {Lekner}},\ }\href {\doibase 10.1103/PhysRev.158.130} {\bibfield  {journal}
  {\bibinfo  {journal} {Phys. Rev.}\ }\textbf {\bibinfo {volume} {158}},\
  \bibinfo {pages} {130} (\bibinfo {year} {1967})}\BibitemShut {NoStop}%
\bibitem [{\citenamefont {Lamp}(1989)}]{lamp1989}%
  \BibitemOpen
  \bibfield  {author} {\bibinfo {author} {\bibfnamefont {P.}~\bibnamefont
  {Lamp}},\ }\emph {\bibinfo {title} {{Untersuchungen zur photoelektrischen
  Injektion von Elektronen in fluessiges Argon}}},\ \href@noop {} {Ph.D.
  thesis},\ \bibinfo  {school} {Technische Universit\H at}, \bibinfo {address}
  {Munich, Germany} (\bibinfo {year} {1989})\BibitemShut {NoStop}%
\bibitem [{\citenamefont {Eibl}, \citenamefont {Lamp},\ and\ \citenamefont
  {Buschhorn}(1990)}]{lamp1990}%
  \BibitemOpen
  \bibfield  {author} {\bibinfo {author} {\bibfnamefont {E.}~\bibnamefont
  {Eibl}}, \bibinfo {author} {\bibfnamefont {P.}~\bibnamefont {Lamp}}, \ and\
  \bibinfo {author} {\bibfnamefont {G.}~\bibnamefont {Buschhorn}},\ }\href
  {\doibase 10.1103/PhysRevB.42.4356} {\bibfield  {journal} {\bibinfo
  {journal} {Physical Review B}\ }\textbf {\bibinfo {volume} {42}},\ \bibinfo
  {pages} {4356} (\bibinfo {year} {1990})}\BibitemShut {NoStop}%
\bibitem [{\citenamefont {Torzo}(1990)}]{torzo1990}%
  \BibitemOpen
  \bibfield  {author} {\bibinfo {author} {\bibfnamefont {G.}~\bibnamefont
  {Torzo}},\ }\href@noop {} {\bibfield  {journal} {\bibinfo  {journal} {Rev.
  Sci. Instrum.}\ }\textbf {\bibinfo {volume} {61}},\ \bibinfo {pages} {1162}
  (\bibinfo {year} {1990})}\BibitemShut {NoStop}%
\bibitem [{\citenamefont {Ch.Tegeler}, \citenamefont {Span},\ and\
  \citenamefont {Wagner}(1997)}]{Tegeler1997}%
  \BibitemOpen
  \bibfield  {author} {\bibinfo {author} {\bibnamefont {Ch.Tegeler}}, \bibinfo
  {author} {\bibfnamefont {R.}~\bibnamefont {Span}}, \ and\ \bibinfo {author}
  {\bibfnamefont {W.}~\bibnamefont {Wagner}},\ }\href@noop {} {\emph {\bibinfo
  {title} {Eine neue Fundamentalgleichung f{\"u}r das fluide Zustandsgebiet von
  Argon f{\"u}r Temperaturen von der Schmelzlinie bis 700 K und Dr{\"u}cke bis
  1000 MPa}}},\ Vol.~\bibinfo {volume} {3}\ (\bibinfo  {publisher}
  {VDI-Verl.},\ \bibinfo {address} {Duesseldorf},\ \bibinfo {year}
  {1997})\BibitemShut {NoStop}%
\bibitem [{\citenamefont {Borghesani}\ and\ \citenamefont
  {Santini}(1990{\natexlab{b}})}]{borg1990b}%
  \BibitemOpen
  \bibfield  {author} {\bibinfo {author} {\bibfnamefont {A.~F.}\ \bibnamefont
  {Borghesani}}\ and\ \bibinfo {author} {\bibfnamefont {M.}~\bibnamefont
  {Santini}},\ }\href@noop {} {\bibfield  {journal} {\bibinfo  {journal} {Meas.
  Sci. Technol.}\ }\textbf {\bibinfo {volume} {1}},\ \bibinfo {pages} {939}
  (\bibinfo {year} {1990}{\natexlab{b}})}\BibitemShut {NoStop}%
\bibitem [{\citenamefont {Kojima}, \citenamefont {Sakai},\ and\ \citenamefont
  {Tagashira}(1987)}]{sakai1987}%
  \BibitemOpen
  \bibfield  {author} {\bibinfo {author} {\bibfnamefont {H.}~\bibnamefont
  {Kojima}}, \bibinfo {author} {\bibfnamefont {Y.}~\bibnamefont {Sakai}}, \
  and\ \bibinfo {author} {\bibfnamefont {H.}~\bibnamefont {Tagashira}},\ }\href
  {\doibase 10.1016/0009-2614(87)80501-8} {\bibfield  {journal} {\bibinfo
  {journal} {Chem. Phys. Lett.}\ }\textbf {\bibinfo {volume} {140}},\ \bibinfo
  {pages} {631} (\bibinfo {year} {1987})}\BibitemShut {NoStop}%
\bibitem [{\citenamefont {Nakamura}, \citenamefont {Sakai},\ and\ \citenamefont
  {Tagashira}(1991)}]{sakai1991}%
  \BibitemOpen
  \bibfield  {author} {\bibinfo {author} {\bibfnamefont {S.}~\bibnamefont
  {Nakamura}}, \bibinfo {author} {\bibfnamefont {Y.}~\bibnamefont {Sakai}}, \
  and\ \bibinfo {author} {\bibfnamefont {H.}~\bibnamefont {Tagashira}},\
  }\href@noop {} {\bibfield  {journal} {\bibinfo  {journal} {J. Phys. D:
  Applied Physics}\ }\textbf {\bibinfo {volume} {24}},\ \bibinfo {pages} {360}
  (\bibinfo {year} {1991})}\BibitemShut {NoStop}%
\bibitem [{\citenamefont {Borghesani}, \citenamefont {Neri},\ and\
  \citenamefont {Barbarotto}(1997)}]{cpl}%
  \BibitemOpen
  \bibfield  {author} {\bibinfo {author} {\bibfnamefont {A.}~\bibnamefont
  {Borghesani}}, \bibinfo {author} {\bibfnamefont {D.}~\bibnamefont {Neri}}, \
  and\ \bibinfo {author} {\bibfnamefont {A.}~\bibnamefont {Barbarotto}},\
  }\href {\doibase 10.1016/S0009-2614(97)00066-3} {\bibfield  {journal}
  {\bibinfo  {journal} {Chem. Phys. Lett.}\ }\textbf {\bibinfo {volume}
  {267}},\ \bibinfo {pages} {116} (\bibinfo {year} {1997})}\BibitemShut
  {NoStop}%
\bibitem [{\citenamefont {Springett}, \citenamefont {Jortner},\ and\
  \citenamefont {Cohen}(1968)}]{SJC}%
  \BibitemOpen
  \bibfield  {author} {\bibinfo {author} {\bibfnamefont {B.~E.}\ \bibnamefont
  {Springett}}, \bibinfo {author} {\bibfnamefont {J.}~\bibnamefont {Jortner}},
  \ and\ \bibinfo {author} {\bibfnamefont {M.~H.}\ \bibnamefont {Cohen}},\
  }\href@noop {} {\bibfield  {journal} {\bibinfo  {journal} {J. Chem. Phys.}\
  }\textbf {\bibinfo {volume} {48}},\ \bibinfo {pages} {2720} (\bibinfo {year}
  {1968})}\BibitemShut {NoStop}%
\bibitem [{\citenamefont {Reininger}\ \emph {et~al.}(1983)\citenamefont
  {Reininger}, \citenamefont {Asaf}, \citenamefont {Steinberger},\ and\
  \citenamefont {Basak}}]{reininger1983}%
  \BibitemOpen
  \bibfield  {author} {\bibinfo {author} {\bibfnamefont {R.}~\bibnamefont
  {Reininger}}, \bibinfo {author} {\bibfnamefont {U.}~\bibnamefont {Asaf}},
  \bibinfo {author} {\bibfnamefont {I.~T.}\ \bibnamefont {Steinberger}}, \ and\
  \bibinfo {author} {\bibfnamefont {S.}~\bibnamefont {Basak}},\ }\href
  {\doibase 10.1103/PhysRevB.28.4426} {\bibfield  {journal} {\bibinfo
  {journal} {Phys. Rev. B}\ }\textbf {\bibinfo {volume} {28}},\ \bibinfo
  {pages} {4426} (\bibinfo {year} {1983})}\BibitemShut {NoStop}%
\bibitem [{\citenamefont {Plenkiewicz}\ \emph {et~al.}(1986)\citenamefont
  {Plenkiewicz}, \citenamefont {Jay-Gerin}, \citenamefont {Plenkiewicz},\ and\
  \citenamefont {Bachelet}}]{bachelet1986}%
  \BibitemOpen
  \bibfield  {author} {\bibinfo {author} {\bibfnamefont {B.}~\bibnamefont
  {Plenkiewicz}}, \bibinfo {author} {\bibfnamefont {J.-P.}\ \bibnamefont
  {Jay-Gerin}}, \bibinfo {author} {\bibfnamefont {P.}~\bibnamefont
  {Plenkiewicz}}, \ and\ \bibinfo {author} {\bibfnamefont {G.~B.}\ \bibnamefont
  {Bachelet}},\ }\href@noop {} {\bibfield  {journal} {\bibinfo  {journal} {Eur.
  Phys. Lett.}\ }\textbf {\bibinfo {volume} {1}},\ \bibinfo {pages} {455}
  (\bibinfo {year} {1986})}\BibitemShut {NoStop}%
\bibitem [{\citenamefont {Fowler}(1931)}]{Fowler1931}%
  \BibitemOpen
  \bibfield  {author} {\bibinfo {author} {\bibfnamefont {R.~H.}\ \bibnamefont
  {Fowler}},\ }\href {\doibase 10.1103/PhysRev.38.45} {\bibfield  {journal}
  {\bibinfo  {journal} {Phys. Rev.}\ }\textbf {\bibinfo {volume} {38}},\
  \bibinfo {pages} {45} (\bibinfo {year} {1931})}\BibitemShut {NoStop}%
\bibitem [{\citenamefont {DuBridge}(1932)}]{dubridge1932}%
  \BibitemOpen
  \bibfield  {author} {\bibinfo {author} {\bibfnamefont {L.~A.}\ \bibnamefont
  {DuBridge}},\ }\href {\doibase 10.1103/PhysRev.39.108} {\bibfield  {journal}
  {\bibinfo  {journal} {Phys. Rev.}\ }\textbf {\bibinfo {volume} {39}},\
  \bibinfo {pages} {108} (\bibinfo {year} {1932})}\BibitemShut {NoStop}%
\bibitem [{\citenamefont {DuBridge}(1933)}]{dubridge1933}%
  \BibitemOpen
  \bibfield  {author} {\bibinfo {author} {\bibfnamefont {L.~A.}\ \bibnamefont
  {DuBridge}},\ }\href {\doibase 10.1103/PhysRev.43.727} {\bibfield  {journal}
  {\bibinfo  {journal} {Phys. Rev.}\ }\textbf {\bibinfo {volume} {43}},\
  \bibinfo {pages} {727} (\bibinfo {year} {1933})}\BibitemShut {NoStop}%
\bibitem [{\citenamefont {Jackson}(1999)}]{jackson}%
  \BibitemOpen
  \bibfield  {author} {\bibinfo {author} {\bibfnamefont {J.~D.}\ \bibnamefont
  {Jackson}},\ }\href@noop {} {\emph {\bibinfo {title} {Classical
  Electrodynamics}}}\ (\bibinfo  {publisher} {Wiley},\ \bibinfo {address} {New
  York},\ \bibinfo {year} {1999})\BibitemShut {NoStop}%
\bibitem [{\citenamefont {Maitland}\ \emph {et~al.}(1981)\citenamefont
  {Maitland}, \citenamefont {Rigby}, \citenamefont {Smith},\ and\ \citenamefont
  {Wakeham}}]{maitland}%
  \BibitemOpen
  \bibfield  {author} {\bibinfo {author} {\bibfnamefont {G.~C.}\ \bibnamefont
  {Maitland}}, \bibinfo {author} {\bibfnamefont {M.}~\bibnamefont {Rigby}},
  \bibinfo {author} {\bibfnamefont {E.~B.}\ \bibnamefont {Smith}}, \ and\
  \bibinfo {author} {\bibfnamefont {W.~A.}\ \bibnamefont {Wakeham}},\
  }\href@noop {} {\emph {\bibinfo {title} {{Intermolecular forces. Their origin
  and determination}}}}\ (\bibinfo  {publisher} {Clarendon},\ \bibinfo
  {address} {Oxford},\ \bibinfo {year} {1981})\BibitemShut {NoStop}%
\bibitem [{\citenamefont {Onn}\ and\ \citenamefont {Silver}(1969)}]{onn1969}%
  \BibitemOpen
  \bibfield  {author} {\bibinfo {author} {\bibfnamefont {D.~G.}\ \bibnamefont
  {Onn}}\ and\ \bibinfo {author} {\bibfnamefont {M.}~\bibnamefont {Silver}},\
  }\href@noop {} {\bibfield  {journal} {\bibinfo  {journal} {Phys. Rev.}\
  }\textbf {\bibinfo {volume} {183}},\ \bibinfo {pages} {295} (\bibinfo {year}
  {1969})}\BibitemShut {NoStop}%
\bibitem [{\citenamefont {Allen}, \citenamefont {Kuntz},\ and\ \citenamefont
  {Schmidt}(1984)}]{allen1984}%
  \BibitemOpen
  \bibfield  {author} {\bibinfo {author} {\bibfnamefont {A.~O.}\ \bibnamefont
  {Allen}}, \bibinfo {author} {\bibfnamefont {P.~J.}\ \bibnamefont {Kuntz}}, \
  and\ \bibinfo {author} {\bibfnamefont {W.~F.}\ \bibnamefont {Schmidt}},\
  }\href@noop {} {\bibfield  {journal} {\bibinfo  {journal} {J. Phys. Chem.}\
  }\textbf {\bibinfo {volume} {88}},\ \bibinfo {pages} {3718} (\bibinfo {year}
  {1984})}\BibitemShut {NoStop}%
\bibitem [{\citenamefont {Thomson}\ and\ \citenamefont
  {Thomson}(1928)}]{thomson1928}%
  \BibitemOpen
  \bibfield  {author} {\bibinfo {author} {\bibfnamefont {J.~J.}\ \bibnamefont
  {Thomson}}\ and\ \bibinfo {author} {\bibfnamefont {G.~P.}\ \bibnamefont
  {Thomson}},\ }\href@noop {} {\emph {\bibinfo {title} {Conduction of
  electricity through gases}}}\ (\bibinfo  {publisher} {Cambridge University
  Press},\ \bibinfo {address} {Cambridge},\ \bibinfo {year} {1928})\BibitemShut
  {NoStop}%
\bibitem [{\citenamefont {Loeb}(1955)}]{loeb1955}%
  \BibitemOpen
  \bibfield  {author} {\bibinfo {author} {\bibfnamefont {L.~B.}\ \bibnamefont
  {Loeb}},\ }\href@noop {} {\emph {\bibinfo {title} {Basic processes of gaseous
  electronics}}}\ (\bibinfo  {publisher} {University of California Press},\
  \bibinfo {address} {Berkeley},\ \bibinfo {year} {1955})\BibitemShut {NoStop}%
\bibitem [{\citenamefont {B\'ekirian}(1968)}]{bekirian1968}%
  \BibitemOpen
  \bibfield  {author} {\bibinfo {author} {\bibfnamefont {A.}~\bibnamefont
  {B\'ekirian}},\ }\href@noop {} {\bibfield  {journal} {\bibinfo  {journal} {J.
  Phys. (Paris)}\ }\textbf {\bibinfo {volume} {29}},\ \bibinfo {pages} {434}
  (\bibinfo {year} {1968})}\BibitemShut {NoStop}%
\bibitem [{\citenamefont {Mozumder}(1980)}]{mozumder1980}%
  \BibitemOpen
  \bibfield  {author} {\bibinfo {author} {\bibfnamefont {A.}~\bibnamefont
  {Mozumder}},\ }\href@noop {} {\bibfield  {journal} {\bibinfo  {journal} {J.
  Chem. Phys.}\ }\textbf {\bibinfo {volume} {72}},\ \bibinfo {pages} {6289}
  (\bibinfo {year} {1980})}\BibitemShut {NoStop}%
\bibitem [{\citenamefont {Weyhreter}\ \emph {et~al.}(1988)\citenamefont
  {Weyhreter}, \citenamefont {Barczik}, \citenamefont {Mann},\ and\
  \citenamefont {Linder}}]{wey1988}%
  \BibitemOpen
  \bibfield  {author} {\bibinfo {author} {\bibfnamefont {M.}~\bibnamefont
  {Weyhreter}}, \bibinfo {author} {\bibfnamefont {B.}~\bibnamefont {Barczik}},
  \bibinfo {author} {\bibfnamefont {A.}~\bibnamefont {Mann}}, \ and\ \bibinfo
  {author} {\bibfnamefont {F.}~\bibnamefont {Linder}},\ }\href@noop {}
  {\bibfield  {journal} {\bibinfo  {journal} {Z. Phys. D: Atoms Mol. Clusters}\
  }\textbf {\bibinfo {volume} {7}},\ \bibinfo {pages} {333} (\bibinfo {year}
  {1988})}\BibitemShut {NoStop}%
\bibitem [{\citenamefont {Haddad}\ and\ \citenamefont
  {O'Malley}(1982)}]{haddad1982}%
  \BibitemOpen
  \bibfield  {author} {\bibinfo {author} {\bibfnamefont {G.~N.}\ \bibnamefont
  {Haddad}}\ and\ \bibinfo {author} {\bibfnamefont {T.~F.}\ \bibnamefont
  {O'Malley}},\ }\href@noop {} {\bibfield  {journal} {\bibinfo  {journal}
  {Aust. J. Phys.}\ }\textbf {\bibinfo {volume} {35}},\ \bibinfo {pages} {35}
  (\bibinfo {year} {1982})}\BibitemShut {NoStop}%
\bibitem [{\citenamefont {Levine}\ and\ \citenamefont
  {Sanders}(1967)}]{levine1967}%
  \BibitemOpen
  \bibfield  {author} {\bibinfo {author} {\bibfnamefont {J.~L.}\ \bibnamefont
  {Levine}}\ and\ \bibinfo {author} {\bibfnamefont {T.~M.}\ \bibnamefont
  {Sanders}},\ }\href@noop {} {\bibfield  {journal} {\bibinfo  {journal} {Phys.
  Rev.}\ }\textbf {\bibinfo {volume} {154}},\ \bibinfo {pages} {138} (\bibinfo
  {year} {1967})}\BibitemShut {NoStop}%
\bibitem [{\citenamefont {Borghesani}\ \emph {et~al.}(1985)\citenamefont
  {Borghesani}, \citenamefont {Bruschi}, \citenamefont {Santini},\ and\
  \citenamefont {Torzo}}]{Borghesani1985}%
  \BibitemOpen
  \bibfield  {author} {\bibinfo {author} {\bibfnamefont {A.}~\bibnamefont
  {Borghesani}}, \bibinfo {author} {\bibfnamefont {L.}~\bibnamefont {Bruschi}},
  \bibinfo {author} {\bibfnamefont {M.}~\bibnamefont {Santini}}, \ and\
  \bibinfo {author} {\bibfnamefont {G.}~\bibnamefont {Torzo}},\ }\href
  {\doibase 10.1016/0375-9601(85)90741-8} {\bibfield  {journal} {\bibinfo
  {journal} {Phys. Lett.}\ }\textbf {\bibinfo {volume} {108 A}},\ \bibinfo
  {pages} {255} (\bibinfo {year} {1985})}\BibitemShut {NoStop}%
\bibitem [{\citenamefont {Bartels}(1973)}]{Bartels1973}%
  \BibitemOpen
  \bibfield  {author} {\bibinfo {author} {\bibfnamefont {A.~K.}\ \bibnamefont
  {Bartels}},\ }\href {\doibase 10.1016/0375-9601(73)90842-6} {\bibfield
  {journal} {\bibinfo  {journal} {Phys. Lett. A}\ }\textbf {\bibinfo {volume}
  {44}},\ \bibinfo {pages} {403} (\bibinfo {year} {1973})}\BibitemShut
  {NoStop}%
\bibitem [{\citenamefont {Borghesani}, \citenamefont {Carugno},\ and\
  \citenamefont {Santini}(1991)}]{Borghesani1991}%
  \BibitemOpen
  \bibfield  {author} {\bibinfo {author} {\bibfnamefont {A.~F.}\ \bibnamefont
  {Borghesani}}, \bibinfo {author} {\bibfnamefont {G.}~\bibnamefont {Carugno}},
  \ and\ \bibinfo {author} {\bibfnamefont {M.}~\bibnamefont {Santini}},\ }\href
  {\doibase 10.1109/14.83680} {\bibfield  {journal} {\bibinfo  {journal} {IEEE
  Trans. Dielect. Electr. Insul.}\ }\textbf {\bibinfo {volume} {26}},\ \bibinfo
  {pages} {615} (\bibinfo {year} {1991})}\BibitemShut {NoStop}%
\bibitem [{\citenamefont {Borghesani}\ and\ \citenamefont
  {Lamp}(2003)}]{Borghesani2003}%
  \BibitemOpen
  \bibfield  {author} {\bibinfo {author} {\bibfnamefont {A.~F.}\ \bibnamefont
  {Borghesani}}\ and\ \bibinfo {author} {\bibfnamefont {P.}~\bibnamefont
  {Lamp}},\ }\href@noop {} {\bibfield  {journal} {\bibinfo  {journal} {IEEE
  Trans. Dielect. Electr. Insul.}\ }\textbf {\bibinfo {volume} {10}},\ \bibinfo
  {pages} {977} (\bibinfo {year} {2003})}\BibitemShut {NoStop}%
\bibitem [{\citenamefont {Lax}(1951)}]{lax1951}%
  \BibitemOpen
  \bibfield  {author} {\bibinfo {author} {\bibfnamefont {M.}~\bibnamefont
  {Lax}},\ }\href {\doibase 10.1103/RevModPhys.23.287} {\bibfield  {journal}
  {\bibinfo  {journal} {Rev. Mod. Phys.}\ }\textbf {\bibinfo {volume} {23}},\
  \bibinfo {pages} {287} (\bibinfo {year} {1951})}\BibitemShut {NoStop}%
\bibitem [{\citenamefont {Foldy}(1945)}]{foldy1945}%
  \BibitemOpen
  \bibfield  {author} {\bibinfo {author} {\bibfnamefont {L.~L.}\ \bibnamefont
  {Foldy}},\ }\href {\doibase 10.1103/PhysRev.67.107} {\bibfield  {journal}
  {\bibinfo  {journal} {Phys. Rev.}\ }\textbf {\bibinfo {volume} {67}},\
  \bibinfo {pages} {107} (\bibinfo {year} {1945})}\BibitemShut {NoStop}%
\bibitem [{\citenamefont {Fermi}(1934)}]{fermi1934}%
  \BibitemOpen
  \bibfield  {author} {\bibinfo {author} {\bibfnamefont {E.}~\bibnamefont
  {Fermi}},\ }\href@noop {} {\bibfield  {journal} {\bibinfo  {journal} {Nuovo
  Cimento}\ }\textbf {\bibinfo {volume} {11}},\ \bibinfo {pages} {157}
  (\bibinfo {year} {1934})}\BibitemShut {NoStop}%
\bibitem [{\citenamefont {Wigner}\ and\ \citenamefont {Seitz}(1933)}]{WS}%
  \BibitemOpen
  \bibfield  {author} {\bibinfo {author} {\bibfnamefont {E.}~\bibnamefont
  {Wigner}}\ and\ \bibinfo {author} {\bibfnamefont {F.}~\bibnamefont {Seitz}},\
  }\href {\doibase 10.1103/PhysRev.43.804} {\bibfield  {journal} {\bibinfo
  {journal} {Phys. Rev.}\ }\textbf {\bibinfo {volume} {43}},\ \bibinfo {pages}
  {804} (\bibinfo {year} {1933})}\BibitemShut {NoStop}%
\bibitem [{\citenamefont {Hernandez}\ and\ \citenamefont
  {Martin}(1991)}]{hernandezmartin1991}%
  \BibitemOpen
  \bibfield  {author} {\bibinfo {author} {\bibfnamefont {J.~P.}\ \bibnamefont
  {Hernandez}}\ and\ \bibinfo {author} {\bibfnamefont {L.~W.}\ \bibnamefont
  {Martin}},\ }\href {\doibase 10.1103/PhysRevA.43.4568} {\bibfield  {journal}
  {\bibinfo  {journal} {Phys. Rev. A}\ }\textbf {\bibinfo {volume} {43}},\
  \bibinfo {pages} {4568} (\bibinfo {year} {1991})}\BibitemShut {NoStop}%
\bibitem [{\citenamefont {Lamp}\ and\ \citenamefont
  {Buschhorn}(1994)}]{lamp1994}%
  \BibitemOpen
  \bibfield  {author} {\bibinfo {author} {\bibfnamefont {P.}~\bibnamefont
  {Lamp}}\ and\ \bibinfo {author} {\bibfnamefont {G.}~\bibnamefont
  {Buschhorn}},\ }\href {\doibase 10.1103/PhysRevB.50.16824} {\bibfield
  {journal} {\bibinfo  {journal} {Phys. Rev. B}\ }\textbf {\bibinfo {volume}
  {50}},\ \bibinfo {pages} {16824} (\bibinfo {year} {1994})}\BibitemShut
  {NoStop}%
\bibitem [{\citenamefont {Ascarelli}(1986)}]{ascarelli1986}%
  \BibitemOpen
  \bibfield  {author} {\bibinfo {author} {\bibfnamefont {G.}~\bibnamefont
  {Ascarelli}},\ }\href {\doibase 10.1103/PhysRevB.33.5825} {\bibfield
  {journal} {\bibinfo  {journal} {Phys. Rev. B}\ }\textbf {\bibinfo {volume}
  {33}},\ \bibinfo {pages} {5825} (\bibinfo {year} {1986})}\BibitemShut
  {NoStop}%
\bibitem [{\citenamefont {Atrazhev}\ and\ \citenamefont
  {Iakubov}(1977)}]{atrazhev1977}%
  \BibitemOpen
  \bibfield  {author} {\bibinfo {author} {\bibfnamefont {V.~M.}\ \bibnamefont
  {Atrazhev}}\ and\ \bibinfo {author} {\bibfnamefont {I.~T.}\ \bibnamefont
  {Iakubov}},\ }\href@noop {} {\bibfield  {journal} {\bibinfo  {journal} {J.
  Phys. D: Appl. Phys.}\ }\textbf {\bibinfo {volume} {10}} (\bibinfo {year}
  {1977})}\BibitemShut {NoStop}%
\bibitem [{\citenamefont {Polischuk}(1985)}]{polischuk}%
  \BibitemOpen
  \bibfield  {author} {\bibinfo {author} {\bibfnamefont {A.~Y.}\ \bibnamefont
  {Polischuk}},\ }\href@noop {} {\bibfield  {journal} {\bibinfo  {journal} {J.
  Phys. B: At. Mol. Phys.}\ }\textbf {\bibinfo {volume} {18}} (\bibinfo {year}
  {1985})}\BibitemShut {NoStop}%
\bibitem [{\citenamefont {Cohen}\ and\ \citenamefont
  {Lekner}(1967)}]{cohen1967}%
  \BibitemOpen
  \bibfield  {author} {\bibinfo {author} {\bibfnamefont {M.~H.}\ \bibnamefont
  {Cohen}}\ and\ \bibinfo {author} {\bibfnamefont {J.}~\bibnamefont {Lekner}},\
  }\href {\doibase 10.1103/PhysRev.158.305} {\bibfield  {journal} {\bibinfo
  {journal} {Phys. Rev.}\ }\textbf {\bibinfo {volume} {158}},\ \bibinfo {pages}
  {305} (\bibinfo {year} {1967})}\BibitemShut {NoStop}%
\bibitem [{\citenamefont {Lekner}(1968)}]{lekner1968}%
  \BibitemOpen
  \bibfield  {author} {\bibinfo {author} {\bibfnamefont {J.}~\bibnamefont
  {Lekner}},\ }\href@noop {} {\bibfield  {journal} {\bibinfo  {journal} {Phil.
  Mag.}\ }\textbf {\bibinfo {volume} {18}},\ \bibinfo {pages} {1281} (\bibinfo
  {year} {1968})}\BibitemShut {NoStop}%
\bibitem [{\citenamefont {Lekner}\ and\ \citenamefont
  {Bishop}(1973)}]{lekner1973}%
  \BibitemOpen
  \bibfield  {author} {\bibinfo {author} {\bibfnamefont {J.}~\bibnamefont
  {Lekner}}\ and\ \bibinfo {author} {\bibfnamefont {A.~R.}\ \bibnamefont
  {Bishop}},\ }\href@noop {} {\bibfield  {journal} {\bibinfo  {journal} {Phil.
  Mag.}\ }\textbf {\bibinfo {volume} {27}},\ \bibinfo {pages} {297} (\bibinfo
  {year} {1973})}\BibitemShut {NoStop}%
\bibitem [{\citenamefont {Steinberger}\ and\ \citenamefont
  {Zeitak}(1986)}]{steinberger1986}%
  \BibitemOpen
  \bibfield  {author} {\bibinfo {author} {\bibfnamefont {I.~T.}\ \bibnamefont
  {Steinberger}}\ and\ \bibinfo {author} {\bibfnamefont {R.}~\bibnamefont
  {Zeitak}},\ }\href {\doibase 10.1103/PhysRevB.34.3471} {\bibfield  {journal}
  {\bibinfo  {journal} {Phys. Rev. B}\ }\textbf {\bibinfo {volume} {34}},\
  \bibinfo {pages} {3471} (\bibinfo {year} {1986})}\BibitemShut {NoStop}%
\bibitem [{\citenamefont {Wannier}(1966)}]{wannier}%
  \BibitemOpen
  \bibfield  {author} {\bibinfo {author} {\bibfnamefont {G.~H.}\ \bibnamefont
  {Wannier}},\ }\href@noop {} {\emph {\bibinfo {title} {{Statistical
  Physics}}}}\ (\bibinfo  {publisher} {Dover},\ \bibinfo {address} {New York},\
  \bibinfo {year} {1966})\BibitemShut {NoStop}%
\bibitem [{\citenamefont {Johnson}, \citenamefont {Broomall},\ and\
  \citenamefont {Onn}(1979)}]{johnson1979}%
  \BibitemOpen
  \bibfield  {author} {\bibinfo {author} {\bibfnamefont {W.~D.}\ \bibnamefont
  {Johnson}}, \bibinfo {author} {\bibfnamefont {J.~R.}\ \bibnamefont
  {Broomall}}, \ and\ \bibinfo {author} {\bibfnamefont {D.~G.}\ \bibnamefont
  {Onn}},\ }\href@noop {} {\bibfield  {journal} {\bibinfo  {journal} {J. Low
  Temp. Phys.}\ }\textbf {\bibinfo {volume} {35}},\ \bibinfo {pages} {535}
  (\bibinfo {year} {1979})}\BibitemShut {NoStop}%
\bibitem [{\citenamefont {Borghesani}, \citenamefont {Neri},\ and\
  \citenamefont {Barbarotto}(1999)}]{o2m}%
  \BibitemOpen
  \bibfield  {author} {\bibinfo {author} {\bibfnamefont {A.}~\bibnamefont
  {Borghesani}}, \bibinfo {author} {\bibfnamefont {D.}~\bibnamefont {Neri}}, \
  and\ \bibinfo {author} {\bibfnamefont {A.}~\bibnamefont {Barbarotto}},\
  }\href@noop {} {\bibfield  {journal} {\bibinfo  {journal} {Int. J.
  Thermophys.}\ }\textbf {\bibinfo {volume} {20}},\ \bibinfo {pages} {899}
  (\bibinfo {year} {1999})}\BibitemShut {NoStop}%
\bibitem [{\citenamefont {Bird}, \citenamefont {Stewart},\ and\ \citenamefont
  {Lightfoot}(1960)}]{byron}%
  \BibitemOpen
  \bibfield  {author} {\bibinfo {author} {\bibfnamefont {R.~B.}\ \bibnamefont
  {Bird}}, \bibinfo {author} {\bibfnamefont {W.~E.}\ \bibnamefont {Stewart}}, \
  and\ \bibinfo {author} {\bibfnamefont {E.~N.}\ \bibnamefont {Lightfoot}},\
  }\href@noop {} {\emph {\bibinfo {title} {Transport phenomena}}}\ (\bibinfo
  {publisher} {Wiley},\ \bibinfo {address} {New York},\ \bibinfo {year}
  {1960})\BibitemShut {NoStop}%
\bibitem [{\citenamefont {Atkins}(1959)}]{atkins1959}%
  \BibitemOpen
  \bibfield  {author} {\bibinfo {author} {\bibfnamefont {K.~R.}\ \bibnamefont
  {Atkins}},\ }\href@noop {} {\bibfield  {journal} {\bibinfo  {journal} {Phys.
  Rev.}\ }\textbf {\bibinfo {volume} {116}},\ \bibinfo {pages} {1339} (\bibinfo
  {year} {1959})}\BibitemShut {NoStop}%
\bibitem [{\citenamefont {Borghesani}, \citenamefont {Neri},\ and\
  \citenamefont {Santini}(1993)}]{neri1993}%
  \BibitemOpen
  \bibfield  {author} {\bibinfo {author} {\bibfnamefont {A.~F.}\ \bibnamefont
  {Borghesani}}, \bibinfo {author} {\bibfnamefont {D.}~\bibnamefont {Neri}}, \
  and\ \bibinfo {author} {\bibfnamefont {M.}~\bibnamefont {Santini}},\
  }\href@noop {} {\bibfield  {journal} {\bibinfo  {journal} {Phys. Rev. E}\
  }\textbf {\bibinfo {volume} {48}},\ \bibinfo {pages} {1379} (\bibinfo {year}
  {1993})}\BibitemShut {NoStop}%
\bibitem [{\citenamefont {Thoen}, \citenamefont {van Geel},\ and\ \citenamefont
  {van Dael}(1971)}]{thoen1971}%
  \BibitemOpen
  \bibfield  {author} {\bibinfo {author} {\bibfnamefont {J.}~\bibnamefont
  {Thoen}}, \bibinfo {author} {\bibfnamefont {E.}~\bibnamefont {van Geel}}, \
  and\ \bibinfo {author} {\bibfnamefont {W.}~\bibnamefont {van Dael}},\ }\href
  {http://www.sciencedirect.com/science/article/pii/0031891471901959}
  {\bibfield  {journal} {\bibinfo  {journal} {Physica}\ }\textbf {\bibinfo
  {volume} {52}},\ \bibinfo {pages} {205} (\bibinfo {year} {1971})}\BibitemShut
  {NoStop}%
\bibitem [{\citenamefont {Kittel}(2004)}]{kittel}%
  \BibitemOpen
  \bibfield  {author} {\bibinfo {author} {\bibfnamefont {C.}~\bibnamefont
  {Kittel}},\ }\href@noop {} {\emph {\bibinfo {title} {Introduction to solid
  state physics}}}\ (\bibinfo  {publisher} {Wiley},\ \bibinfo {address} {New
  York},\ \bibinfo {year} {2004})\BibitemShut {NoStop}%
\bibitem [{\citenamefont {Guggenheim}(1977)}]{guggenheim}%
  \BibitemOpen
  \bibfield  {author} {\bibinfo {author} {\bibfnamefont {E.~A.}\ \bibnamefont
  {Guggenheim}},\ }\href@noop {} {\emph {\bibinfo {title} {Thermodynamics}}}\
  (\bibinfo  {publisher} {North-Holland},\ \bibinfo {address} {Amsterdam},\
  \bibinfo {year} {1977})\BibitemShut {NoStop}%
\bibitem [{\citenamefont {Chandler}(1987)}]{chandler}%
  \BibitemOpen
  \bibfield  {author} {\bibinfo {author} {\bibfnamefont {D.}~\bibnamefont
  {Chandler}},\ }\href@noop {} {\emph {\bibinfo {title} {Introduction to modern
  statistical physics}}}\ (\bibinfo  {publisher} {Oxford University Press},\
  \bibinfo {address} {Oxford},\ \bibinfo {year} {1987})\BibitemShut {NoStop}%
\bibitem [{\citenamefont {Broomall}, \citenamefont {Johnson},\ and\
  \citenamefont {Onn}(1976)}]{broomall1976}%
  \BibitemOpen
  \bibfield  {author} {\bibinfo {author} {\bibfnamefont {J.~R.}\ \bibnamefont
  {Broomall}}, \bibinfo {author} {\bibfnamefont {W.~D.}\ \bibnamefont
  {Johnson}}, \ and\ \bibinfo {author} {\bibfnamefont {D.~G.}\ \bibnamefont
  {Onn}},\ }\href@noop {} {\bibfield  {journal} {\bibinfo  {journal} {Phys.
  Rev. B}\ }\textbf {\bibinfo {volume} {14}},\ \bibinfo {pages} {2819}
  (\bibinfo {year} {1976})}\BibitemShut {NoStop}%
\end{thebibliography}
\end{document}